\documentclass[12pt]{iopart}

\usepackage{iopams}
\usepackage{graphicx}
\begin{document}

\title{Information geometry, trade-off relations, and generalized Glansdorff-Prigogine criterion for stability}

\author{Sosuke Ito}

\address{$^1$ Universal Biology Institute, The University of Tokyo, 7-3-1 Hongo, Bunkyo-ku, Tokyo 113-0031, Japan\\$^2$JST, PRESTO, 4-1-8 Honcho, Kawaguchi, Saitama, 332-0012, Japan}
\ead{sosuke.ito@ubi.s.u-tokyo.ac.jp}
\vspace{10pt}
\begin{indented}
\item[]July 2021
\end{indented}

\begin{abstract}
We discuss a relationship between information geometry and the Glansdorff-Prigogine criterion for stability. For the linear master equation, we found a relation between the line element and the excess entropy production rate. This relation leads to a new perspective of stability in a nonequilibrium steady-state. We also generalize the Glansdorff-Prigogine criterion for stability based on information geometry.  Our information-geometric criterion for stability works well for the nonlinear master equation, where the Glansdorff-Prigogine criterion for stability does not work well. We derive a trade-off relation among the fluctuation of the observable, the mean change of the observable, and the intrinsic speed. We also derive a novel thermodynamic trade-off relation between the excess entropy production rate and the intrinsic speed. These trade-off relations provide a physical interpretation of our information-geometric criterion for stability. We illustrate our information-geometric criterion for stability by an autocatalytic reaction model, where dynamics are driven by a nonlinear master equation.
\end{abstract}

\maketitle
\section{Introduction}
The behavior of the system around the nonequilibrium steady-state has been well discussed in nonequilibrium thermodynamics. Around 1970, P. Glandorff and I. Prigogine proposed the criterion for stability based on a generalization of the entropy production rate for the steady-state, namely the excess entropy production rate~\cite{Glansdorff1964criterion, Glansdorff1970stability, Glansdorff1971stability}. This criterion is called the Glansdorff-Prigogine criterion for stability. They proposed the positivity of the excess entropy production rate as a simple sufficient condition of the stability in the steady-state. The range of the validity of this criterion has been exposed to criticism sometimes~\cite{Tomita1972remarks, Keizer1974criterion, Glansdorff1974criterion}. Nowadays, this criterion has been regarded as the Lyapunov criterion for a particular system~\cite{Sobrino1975stability, Wilhelm1999what}. For example, in the linear master equation, a relationship between the Lyapunov criterion and the Glansdorff-Prigogine criterion for stability has been discussed based on the linear irreversible thermodynamics~\cite{schlogl1971stability, schnakenberg1976network}. Recently, nonequilibrium thermodynamics for the linear master equation has been intensively studied in terms of stochastic thermodynamics~\cite{sekimoto2010stochastic,seifert2012stochastic}. The idea of excess entropy production rate has been widely used in stochastic thermodynamics~\cite{Hatano2001Steadystate, Komatsu2008Steadystate, Komatsu2008Steadystate2,Maes2014Steadystate}, and the Glansdorff-Prigogine criterion for stability has been recently revisited~\cite{Maes2015revisiting}. While the Glansdorff-Prigogine criterion for stability is reasonable for the linear master equation, the Glansdorff-Prigogine criterion for stability does not work well for the nonlinear master equation~\cite{Keizer1974criterion, Wilhelm1999what}.

In a somewhat different context, the connection between information theory and stochastic thermodynamics has been intensively discussed in the past decade~\cite{parrondo2015thermodynamics, Allahverdyan2009thermodynamic,sagawa2010generalized,ito2013information,hartich2014stochastic,horowitz2014thermodynamics,ito2015maxwell,Hartich2016sensory}. The relation between the entropy production rate and informational quantity has been discussed based on the nonnegativity of the Kullback-Leibler divergence~\cite{kawai2007dissipation,ito2016backward,Ito2018unified}. The Lyapunov candidate function in the Glansdorff-Prigogine criterion for stability is related to the Kullback-Leibler divergence~\cite{schnakenberg1976network}. Because the Kullback-Leibler divergence gives a differential geometry known as information geometry~\cite{Rao1992information, Amari2007infogeo, Kass2011Geometrical, Amari2016infogeo, Ay2017infogeo}, information geometry has been discussed from the viewpoint of equilibrium thermodynamics~\cite{Weinhold1975metric, Ruppeiner1979thermodynamics, Salamon1983thermodynamic, Crooks2007length,Sivak2012optimal,Rotskoff2015optimal,Nakamura2019reconsideration,Aguilera2021unified}, stochastic thermodynamics~\cite{Ito2018unified,Polettini2013nonconvexity,Lahiri2016tradeoff,Rotskoff2017geometric,Ito2018geometry,Ito2018cramerrao, gputa2020tighter,kolchinsky2020work}, and chemical thermodynamics~\cite{Yoshimura2021information,Ashida2020information}. 

In recent years, thermodynamic trade-off relations have been well discussed in terms of thermodynamic uncertainty relations~\cite{horowitz2020thermodynamic,Barato2015uncertainty,Pietzonka2016uncertainty,Gingrich2016dissipation}. The thermodynamic uncertainty relations are based on the mathematical property of the Fisher information matrix~\cite{Ito2018cramerrao,Dechant2018multidimensional,Hasegawa2019uncertainty}. This Fisher information matrix becomes a Riemannian metric of information geometry~\cite{Amari2016infogeo}, and several thermodynamic trade-off relations have been derived from the mathematical property of information geometry~\cite{Ito2018geometry,Ito2018cramerrao, gputa2020tighter,Yoshimura2021information,Ashida2020information, nicholson2020time}. These thermodynamic trade-off relations mainly focus on a trade-off between the fluctuation of observable and the entropy production rate. The concept of thermodynamic trade-off relations is similar to the Glansdorff-Prigogine criterion for stability because the excess entropy production rate decides the stability of the steady-state.

This article clarifies a relationship between the Glansdorff-Prigogine criterion for stability and information geometry and discusses the stability of the steady-state based on thermodynamic trade-off relations. We show that the excess entropy production rate is given by the time derivative of the Lyapunov  candidate function for the linear master equation, and this Lyapunov candidate function is related to the square of the line element in information geometry.
This relation gives a geometric interpretation of the Onsager matrix under the near-equilibrium condition. We discuss the stability of the steady-state in terms of the monotonicity of the Kullback-Leibler divergence. 

We also generalize the Glansdorff-Prigogine criterion for stability based on information geometry. Our information-geometric criterion for stability focus on the intrinsic speed on the manifold of probability simplex. If the intrinsic speed is reducing in time around the steady-state, the steady-state is stable. Our generalized criterion for stability works well even for the nonlinear master equation where the Glansdorff-Prigogine criterion for stability does not work well. We also derive trade-off relations among the fluctuation of the observable, the mean change of the observable, and the intrinsic speed, and a novel thermodynamic trade-off relation between the excess entropy production rate and the intrinsic speed. We discuss a physical interpretation of our information-geometric criterion for stability based on these trade-off relations. The speed of observable decreases around the steady-state if the steady-state is stable. We illustrate the merit of our information-geometric criterion for stability by a model of autocatalytic reaction driven by a nonlinear master equation.

\section{Review on the Gransdorff-Prigoine criterion for stability}
\subsection{The entropy production rate and the Lyapunov candidate function}  
We first explain a relation between the entropy production rate and the Lyapunov candidate function for the relaxation to equilibrium~\cite{schnakenberg1976network}. We start with the master equation,
\begin{eqnarray}
\frac{d p_i}{dt} = \sum_{j} [W_{j \to i} p_j  - W_{i \to j} p_i],
\label{master}
\end{eqnarray}
where $W_{i \to j}$ is the transition rate from the state $i$ to $j$, and $p_i$ is the probability of the state $i$. We introduce the vector notation of the probability $\boldsymbol{p}=\{ {p}_i| i=1, \dots\}$. In this relaxation process, we assume that the transition rate does not depend on time and the detailed balance condition is satisfied. The detail balance condition implies the existence of an equilibrium state $\boldsymbol{p}^{\rm eq}=\{ {p}^{\rm eq}_i| i=1, \dots\}$ that satisfies 
\begin{eqnarray}
W_{j \to i} p^{\rm eq}_j  - W_{i \to j} p^{\rm eq}_i =0,
\label{detailedbalance}
\end{eqnarray}
for any pair $(i,j)$. The equilibrium state is the steady-state, that is 
\begin{eqnarray}
\frac{d\boldsymbol{p}^{\rm eq}}{dt} = \boldsymbol{0}.
\label{eqsteady}
\end{eqnarray}
It implies that the equilibrium state is the fixed point of the master equation.

We now introduce the entropy production rate $\sigma (\boldsymbol{p})$ for the state $\boldsymbol{p}$. We here assume that the transition rate is non-zero $W_{j \to i} \neq 0$ and the Boltzmann constant is set to unity $k_{\rm B}=1$ to simplify the discussion. The entropy production rate $\sigma$ is defined as the product of the force $F_{j \to i} (\boldsymbol{p})$ and the flux $J_{j \to i} (\boldsymbol{p})$,
\begin{eqnarray}
\sigma (\boldsymbol{p}) = \sum_{i,j|j>i} F_{j \to i} (\boldsymbol{p}) J_{j \to i} (\boldsymbol{p}),
\end{eqnarray}
where the flux and the force are defined as
\begin{eqnarray}
J_{j \to i}(\boldsymbol{p}) &= W_{j \to i}p_j - W_{i \to j} p_i, \\
F_{j \to i}(\boldsymbol{p}) &= \ln \frac{W_{j \to i}p_j}{W_{i \to j} p_i},
\end{eqnarray}
respectively. Its nonnegativity $\sigma  (\boldsymbol{p})  \geq 0$ is well known as the second law of thermodynamics. This nonnegativity can be derived from the fact that the flux $J_{j \to i} (\boldsymbol{p})$ and the force $F_{j \to i} (\boldsymbol{p})$ have the same signs.  In equilibrium, the entropy production rate is zero $\sigma  (\boldsymbol{p}^{\rm eq}) = 0$ because of the detailed balance condition Eq.~(\ref{detailedbalance}).

The entropy production rate can be the time derivative of a Lyapunov candidate function if we assume the relaxation to equilibrium. In this relaxation process, the entropy production rate can be rewritten as
\begin{eqnarray}
\sigma  (\boldsymbol{p})  &= \sum_{i,j|j>i} (W_{j \to i}p_j - W_{i \to j} p_i) \left( \ln \frac{p_j }{ p_j^{\rm eq}} -  \ln \frac{p_i }{ p_i^{\rm eq}} \right) \nonumber \\
&=- \sum_{i,j} (W_{j \to i}p_j - W_{i \to j} p_i) \ln \frac{p_i }{ p_i^{\rm eq}}  \nonumber \\
&=- \sum_{i} \frac{dp_i}{dt} \ln \frac{p_i }{ p_i^{\rm eq}} \nonumber \\
&= - \frac{d}{dt} \left( \sum_{i} p_i \ln \frac{p_i }{ p_i^{\rm eq}} \right) \nonumber \\
&:=-\frac{d}{dt} D_{\rm KL}(\boldsymbol{p} ||\boldsymbol{p}^{\rm eq}).
\end{eqnarray}
where we used the detailed balance condition Eq.~(\ref{detailedbalance}), the master equation Eq.~(\ref{master}), the equilibrium property Eq.~(\ref{eqsteady}), and the normalization of the probability $d(\sum_i p_i)/dt =0$. The quantity 
\begin{eqnarray}
D_{\rm KL}(\boldsymbol{p} ||\boldsymbol{p}^{\rm eq}) = \sum_i p_i \ln \frac{p_i}{p^{\rm eq}_i}
\end{eqnarray}
is the Kullback-Leibler divergence between two probabilities $\boldsymbol{p}$ and $\boldsymbol{p}^{\rm eq}$. This quantity is nonnegative and zero if and only if $\boldsymbol{p} =\boldsymbol{p}^{\rm eq}$. Because the equilibrium state $\boldsymbol{p}^{\rm eq}$ is the fixed point of the master equation, the Kullback-Leibler divergence $D_{\rm KL}(\boldsymbol{p} ||\boldsymbol{p}^{\rm eq})$ can be a Lyapunov candidate function. In terms of the Lyapunov criterion, the second law of thermodynamics provides the condition of stability 
\begin{eqnarray}
\frac{d}{dt} D_{\rm KL}(\boldsymbol{p} ||\boldsymbol{p}^{\rm eq}) \leq 0.
\end{eqnarray}
It implies that the equilibrium state is always stable.

\subsection{The excess entropy production rate and the Glansdorff-Prigogine criterion for stability}  
We next revisit the Glansdorff-Prigogine criterion for stability~\cite{schnakenberg1976network}. We consider a steady-state $\boldsymbol{\bar{p}}=\{ \bar{p}_i| i=1, \dots\}$ as a fixed point of the master equation, that is 
\begin{eqnarray}
\frac{d \boldsymbol{\bar{p}}}{dt} = \boldsymbol{0}.
\label{stationary}
\end{eqnarray}
 From the master equation Eq.~(\ref{master}), this condition implies $\sum_{j} {J}_{j \to i} (\boldsymbol{\bar{p}})=0$ for any $i$.
The Glansdorff-Prigogine criterion denotes the stability of a fixed point $\boldsymbol{\bar{p}}$ based on the generalization of the entropy production rate, that is the excess entropy production rate. To define the excess entropy production rate, we introduce the excess flux and the excess force defined as
\begin{eqnarray}
\delta{J}_{j \to i}(\boldsymbol{p} || \bar{\boldsymbol{p}}) &= {J}_{j \to i}(\boldsymbol{p}) - {J}_{j \to i} (\bar{\boldsymbol{p}}),
\\
\delta{F}_{j \to i}(\boldsymbol{p} || \bar{\boldsymbol{p}}) &= {F}_{j \to i} (\boldsymbol{p}) -{F}_{j \to i} (\bar{\boldsymbol{p}}),
\end{eqnarray}
respectively. The excess entropy production rate $\delta^2 \sigma (\boldsymbol{p} || \bar{\boldsymbol{p}})$ is given by the product of the excess flux and the excess force,
\begin{eqnarray}
\delta^2 \sigma  (\boldsymbol{p} || \bar{\boldsymbol{p}})= \sum_{i,j|i>j} \delta F_{i \to j} (\boldsymbol{p} || \bar{\boldsymbol{p}}) \delta J_{i \to j} (\boldsymbol{p} || \bar{\boldsymbol{p}}).
\end{eqnarray}
If the steady-state is given by the equilibrium distribution $\bar{\boldsymbol{p}} ={\boldsymbol{p}}^{\rm eq}$, the detailed balance condition ${J}_{j \to i}(\bar{\boldsymbol{p}})={J}_{j \to i}({\boldsymbol{p}}^{\rm eq})=0$ (and ${F}_{j \to i}(\bar{\boldsymbol{p}})={F}_{j \to i}({\boldsymbol{p}}^{\rm eq})=0$) holds, and we obtain $\delta^2 \sigma  (\boldsymbol{p} || {\boldsymbol{p}}^{\rm eq}) =\sigma (\boldsymbol{p} )$. Thus, this excess entropy production rate is a generalization of the entropy production rate where the detailed balance condition is violated.

In the Glansdorff-Prigogine criterion for stability~\cite{Glansdorff1970stability, Glansdorff1971stability}, we discuss the sign of this excess entropy production rate to detect the stability of the steady-state $\bar{\boldsymbol{p}}$. If it is non-negative
\begin{eqnarray}
\delta^2 \sigma (\boldsymbol{p} || \bar{\boldsymbol{p}}) \geq 0 \: \:\: \:\: \:\: \:\: \:({\rm stable}),
\end{eqnarray}
the steady-state $\bar{\boldsymbol{p}}$ is stable. If it is negative
\begin{eqnarray}
\delta^2 \sigma (\boldsymbol{p} || \bar{\boldsymbol{p}}) < 0 \: \:\: \:\: \: ({\rm unstable}),
\label{unstableGP}
\end{eqnarray}
the steady-state $\bar{\boldsymbol{p}}$ is unstable. 

We discuss a relationship between the Glansdorff-Prigogine criterion for stability and the Lyapunov criterion in the case of the linear master equation, where the transition rate does not depend on the probability $dW_{i \to j}/d{p_k} =0$. To discuss the nonnegativity of the excess entropy production rate, we introduce the difference between the probability distribution and the steady-state distribution
\begin{eqnarray}
\delta p_i  = p_i - \bar{p}_i.
\end{eqnarray}
Up to the order $\delta p_i$, the excess flux and the excess force are calculated as
\begin{eqnarray}
\delta{J}_{j \to i}(\boldsymbol{p} || \bar{\boldsymbol{p}})  &= W_{j \to i} \delta{p}_j - W_{i \to j} \delta{p}_i ,
\label{fluxdelta}
\\
\delta{F}_{j \to i}(\boldsymbol{p} || \bar{\boldsymbol{p}})  &= \frac{\delta p_j }{\bar{p}_j} -\frac{\delta p_i}{\bar{p}_i} + O(|\delta \boldsymbol{p}|^2),
\label{forcedelta}
\end{eqnarray}
respectively, where the order $O(|\delta \boldsymbol{p}|^2)$ is the Landau notation. We find that the time evolution of $\delta p_i$ is given by the sum of the excess fluxes
\begin{eqnarray}
\frac{d }{dt} \delta {p}_i =\sum_{j} \delta {J}_{j \to i} (\boldsymbol{p} || \bar{\boldsymbol{p}}) .
\label{masterdelta}
\end{eqnarray}
From Eqs. (\ref{fluxdelta}), (\ref{forcedelta}), and (\ref{masterdelta}), this excess entropy production rate is calculated as 
\begin{eqnarray}
\delta^2 \sigma   (\boldsymbol{p} || \bar{\boldsymbol{p}}) &  = - \sum_{i,j} \delta {J}_{j \to i}  (\boldsymbol{p} || \bar{\boldsymbol{p}}) \left[\frac{\delta p_i }{\bar{p}_i} \right] \nonumber \\
&  = - \frac{1}{2} \frac{d }{dt}  \left[\sum_{i}\frac{(\delta {p}_i  )^2}{\bar{p}_i} \right] \nonumber \\
&  := - \frac{d }{dt}  \delta^2 \mathcal{L}   (\boldsymbol{p} || \bar{\boldsymbol{p}}).
\label{excesslyapunov}
\end{eqnarray}
The quantity 
\begin{eqnarray}
\delta^2 \mathcal{L}   (\boldsymbol{p} || \bar{\boldsymbol{p}}) = \frac{1}{2} \sum_{i}\frac{(\delta {p}_i )^2}{\bar{p}_i},
\label{Lyapunov}
\end{eqnarray}
is the Lyapunov candidate function because it is nonnegative $\delta^2 \mathcal{L} (\boldsymbol{p} || \bar{\boldsymbol{p}}) \geq 0$, and zero if and only if the probability $\boldsymbol{p}$ is the fixed point $\bar{\boldsymbol{p}}$. Then, the Glansdorff-Prigogine criterion for stability can be considered as the Lyapunov criterion~\cite{Sobrino1975stability} as follows
\begin{eqnarray}
\delta^2 \sigma (\boldsymbol{p} || \bar{\boldsymbol{p}}) \geq 0  \Leftrightarrow \frac{d}{dt}\delta^2 \mathcal{L} &\leq 0 \: \:\: \:\: \: ({\rm stable}), \\
\delta^2 \sigma (\boldsymbol{p} || \bar{\boldsymbol{p}}) < 0  \Leftrightarrow \frac{d}{dt}\delta^2 \mathcal{L} & > 0 \: \:\: \:\: \: ({\rm unstable}).
\end{eqnarray}

We compare this Lyapunov candidate function with the Kullback-Leibler divergence. The Kullback-Leibler divergence between $\boldsymbol{p}$ and $\bar{\boldsymbol{p}}$ is equal to this Lyapunov candidate function $\delta^2 \mathcal{L} (\boldsymbol{p} || \bar{\boldsymbol{p}})$ up to the second order of $\delta {p}_i$,
\begin{eqnarray}
D_{\rm KL} (\boldsymbol{p} || \bar{\boldsymbol{p}}) &= \sum_i p_i \ln \frac{p_i}{\bar{p}_i} \nonumber \\
&= -\sum_i p_i \ln \left( 1 - \frac{\delta p_i}{p_i} \right) \\
&=\delta^2 \mathcal{L}   (\boldsymbol{p} || \bar{\boldsymbol{p}}) + O(|\delta \boldsymbol{p}|^3),
\end{eqnarray}
where we used the normalization of the probability $\sum_i \delta p_i =0$. Thus, this Lyapunov candidate function $\delta^2 \mathcal{L} (\boldsymbol{p} || \bar{\boldsymbol{p}})$  can be regarded as the Kullback-Leibler divergence if $\delta {p}_i$ is small.

\subsection{The stability for the linear master equation and the monotonicity of the Kullback-Leibler divergence}
We discuss the stability for the linear master equation based on the monotonicity of the Kullback-Leibler divergence. We now consider the situation that the transition rate does not depend on time $dW_{i \to j}/dt =0$. The difference equation corresponding to the master equation Eq.~(\ref{master}) is given by
\begin{eqnarray}
{p}_i (t+ \Delta t) &= \sum_j \left(W_{j \to i} \Delta t+ \delta_{ij}\left(1 - \sum_{k} W_{i \to k} \Delta t \right)\right) {p}_j (t) \nonumber \\
&= \sum_j \mathcal{T}^{\Delta t}_{ij} {p}_j (t),
\end{eqnarray} 
where the transition matrix $\mathcal{T}^{\Delta t}_{ij}$ is defined as
\begin{eqnarray}
\mathcal{T}^{\Delta t}_{ij} = W_{j \to i} \Delta t+ \delta_{ij} \left(1 - \sum_{k} W_{i \to k} \Delta t \right).
\end{eqnarray} 
In the vector notation, we write the above difference equation as $\boldsymbol{p}(t+ \Delta t) = \mathcal{T}^{\Delta t} \boldsymbol{p} (t)$. We easily check that the steady-state distribution $\bar{\boldsymbol{p}}$ satisfies $\bar{\boldsymbol{p}} = \mathcal{T}^{\Delta t} \bar{\boldsymbol{p}}$. By using this transition matrix $\mathcal{T}^{\Delta t}_{ij}$, the excess entropy production rate around the steady-state $\boldsymbol{p} (t) \simeq \bar{\boldsymbol{p}}$ is calculated as
\begin{eqnarray}
\delta^2 \sigma   (\boldsymbol{p} (t) || \bar{\boldsymbol{p}})  &= - \frac{d}{dt} \delta^2 \mathcal{L}   (\boldsymbol{p}(t) || \bar{\boldsymbol{p}}) \nonumber \\
&\simeq - \lim_{\Delta t \to 0} \frac{D_{\rm KL} (\boldsymbol{p}(t+\Delta t)|| \bar{\boldsymbol{p}}) - D_{\rm KL} (\boldsymbol{p}(t)|| \bar{\boldsymbol{p}})}{\Delta t} \nonumber \\
&= \lim_{\Delta t \to 0} \frac{ D_{\rm KL} (\boldsymbol{p}(t)|| \bar{\boldsymbol{p}}) - D_{\rm KL} (\mathcal{T}^{\Delta t} \boldsymbol{p}(t)|| \mathcal{T}^{\Delta t} \bar{\boldsymbol{p}})}{\Delta t}.
\end{eqnarray} 
Because the monotonicity of the Kullback-Leibler divergence~\cite{sekimoto2010stochastic} 
\begin{eqnarray}
D_{\rm KL} (\mathcal{T}^{\Delta t} \boldsymbol{p}(t)|| \mathcal{T}^{\Delta t} \bar{\boldsymbol{p}})  \leq D_{\rm KL} (\boldsymbol{p}(t)|| \bar{\boldsymbol{p}})
\end{eqnarray} 
holds, we obtain the nonnegativity of the excess entropy production rate
\begin{eqnarray}
\delta^2 \sigma   (\boldsymbol{p} (t) || \bar{\boldsymbol{p}})  \geq 0,
\end{eqnarray} 
which implies that the steady-state $\bar{\boldsymbol{p}}$ is stable from the viewpoint of the Glansdorff-Prigogine criterion for stability. This fact can be regarded as the principle of minimum entropy production.

We remark that the monotonicity can be violated in the case of the nonlinear master equation. In the case of the nonlinear master equation, the transition matrix generally depends on the probability $\mathcal{T}^{\Delta t}(\boldsymbol{p})$, and the steady-state distribution is given by $\bar{\boldsymbol{p}} = \mathcal{T}^{\Delta t}(\bar{\boldsymbol{p}}) \bar{\boldsymbol{p}}$. Thus, the quantity
\begin{eqnarray}
D_{\rm KL} (\mathcal{T}^{\Delta t}(\boldsymbol{p}(t)) \boldsymbol{p}(t)|| \mathcal{T}^{\Delta t}(\bar{\boldsymbol{p}})\bar{\boldsymbol{p}}) - D_{\rm KL} (\boldsymbol{p}(t)|| \bar{\boldsymbol{p}})
\end{eqnarray} 
is not always nonnegative because the transition matrix $\mathcal{T}^{\Delta t}(\boldsymbol{p}(t))$ is different from the transition matrix $\mathcal{T}^{\Delta t}(\bar{\boldsymbol{p}})$. 

\section{Information geometry and the Glansdorff-Prigogine criterion for stability} 
\subsection{Information geometry and the Lyapunov candidate function}
We here discuss a relationship between the Glansdorff-Prigogine criterion for stability and information geometry. Information geometry is a differential geometry where the Riemannian metric tensor is given by the Fisher information matrix
\begin{eqnarray}
g_{\mu \nu} (\boldsymbol{p}) =\mathbb{E}_{\boldsymbol{p}}[ (\partial_{\theta_{\mu}} \ln p)(\partial_{\theta_{\nu}} \ln p)],
\end{eqnarray}
for the set of parameters $\boldsymbol{\theta}= \{\theta_1, \dots, \theta_N \}$~\cite{Amari2016infogeo}, where $\mathbb{E}_{\boldsymbol{p}}[f] = \sum_{i} p_i f_i$ is the expected value of the function $f_i$.
 The square of the line element $ds^2  (\boldsymbol{p})$ is given by
\begin{eqnarray}
ds^{2} (\boldsymbol{p})&= \sum_{\mu, \nu} g_{\mu \nu} d\theta_{\mu} d\theta_{\nu}.
\end{eqnarray}
The square of the line element does not depend on the choice of parameters because we obtain the following identity
\begin{eqnarray}
ds^{2} (\boldsymbol{p})&= \mathbb{E}_{\boldsymbol{p}}[ (\sum_{\mu} [\partial_{\theta_{\mu}} \ln p ]d \theta_{\mu})(\sum_{\nu}[ \partial_{\theta_{\nu}} \ln p] d \theta_{\nu})] \nonumber \\ 
&= \mathbb{E}_{\boldsymbol{p}}[ (d \ln p)^2 ] \nonumber \\
&= \sum_i \frac{(dp_i)^2}{p_i}.
\end{eqnarray}
 Thus, the square of the line element can be regarded as the Hessian of the Kullback-Leibler divergence
\begin{eqnarray}
ds^2 (\boldsymbol{p})&= 2D_{\rm KL}(\boldsymbol{p} +d\boldsymbol{p}|| \boldsymbol{p})  + O(|d\boldsymbol{p}|^3).
\end{eqnarray}
Thus, the Lyapunov candidate function $\delta^2 \mathcal{L} (\boldsymbol{p} || \bar{\boldsymbol{p}}) $ can be regarded as the square of the line element around the steady-state $\delta \boldsymbol{p}  = d\boldsymbol{p}$ as follows,
\begin{eqnarray}
\delta^2 \mathcal{L}  (\boldsymbol{p} || \bar{\boldsymbol{p}})= D_{\rm KL} (\bar{\boldsymbol{p}} + d\boldsymbol{p}|| \bar{\boldsymbol{p}})  + O(|d\boldsymbol{p}|^3) = \frac{1}{2} ds^2 (\bar{\boldsymbol{p}})  + O(|d\boldsymbol{p}|^3).
\label{relationship}
\end{eqnarray} 
This fact implies that the Glansdorff-Prigogine criterion for stability is the Lyapunov criterion on information-geometric manifold. 

\subsection{The Fisher metric and the Onsager matrix under the near-equilibrium condition}
 We introduce the setup of linear irreversible thermodynamics~\cite{schnakenberg1976network}, which explains the behavior of the system under the near-equilibrium condition. We now start with the equilibrium distribution $\boldsymbol{p}^{\rm eq}$. The detailed balance condition
\begin{eqnarray}
W^{\rm eq}_{j \to i}{p}_j^{\rm eq} =W^{\rm eq}_{i \to j}{p}_i^{\rm eq},
\end{eqnarray}
holds for any pair $(i,j)$, where $W^{\rm eq}_{j \to i}$ is the transition rate for the equilibrium state. We discuss the transition rate $W_{j \to i} = W^{\rm eq}_{j \to i} + O(\epsilon)$ where the order $O(\epsilon)$ implies the small change of the transition rate. This transition rate $W_{j \to i}$ leads to the steady-state state $\bar{\boldsymbol{p}}$ around the equilibrium state $\boldsymbol{p}^{\rm eq}$, where the steady-state state $\bar{\boldsymbol{p}}$ is given by
\begin{eqnarray}
\sum_{j}[W_{j \to i} \bar{p}_j - W_{i \to j} \bar{p}_i ] =0.
\end{eqnarray} 
 In this setup, the steady flux and the steady force are small, ${J}_{j \to i} (\bar{\boldsymbol{p}}) = O(\epsilon)$ and ${F}_{j \to i} (\bar{\boldsymbol{p}})=O(\epsilon)$. The steady force is proportional to the steady flux up to the order $O(\epsilon^2)$,
\begin{eqnarray}
{F}_{j \to i} (\bar{\boldsymbol{p}}) &=&\alpha_{j \to i} {J}_{j \to i} (\bar{\boldsymbol{p}}) + O(\epsilon^2), \\
\alpha_{j \to i}& =&\frac{1}{W^{\rm eq}_{i \to j}{p}_i^{\rm eq}} =\frac{1}{W^{\rm eq}_{j \to i} {p}_j^{\rm eq}}.
\end{eqnarray}
The coefficient $\alpha_{j \to i}$ is determined by parameters in the equilibrium state $W^{\rm eq}_{i \to j}{p}_i^{\rm eq}$. The coefficient is also given by
\begin{eqnarray}
\alpha_{j \to i}& =&\frac{1}{W_{i \to j} \bar{p}_i} +O(\epsilon)=\frac{1}{W_{j \to i} \bar{p}_j} +O(\epsilon),
\label{coefficientorder}
\end{eqnarray}
up to the order $O(\epsilon)$.

We introduce the cycle force $F_{\mu} (\bar{\boldsymbol{p}})$ and the cycle flux $J_{\mu} (\bar{\boldsymbol{p}})$ defined as
\begin{eqnarray} 
F_{\mu} (\bar{\boldsymbol{p}})  =\sum_{i, j | i>j} {F}_{j \to i} (\bar{\boldsymbol{p}})  S_{j \to i}(\mu), \\
J_{j \to i}(\bar{\boldsymbol{p}}) =\sum_{\mu} {J}_{\mu} (\bar{\boldsymbol{p}})  S_{j \to i}(\mu),
\end{eqnarray}
where $S_{j \to i}(\mu)$ is the cycle matrix~\cite{schnakenberg1976network}. The cycle matrix satisfies that $S_{j \to i}(\mu) =-S_{i \to j}(\mu)=1$ if the $\mu$-th cycle includes the edge $j \to i$, and that $S_{j \to i}(\mu)=S_{i \to j}(\mu) =0$ if the $\mu$-th cycle does not include the edge $j \to i$. The $\nu$-th cycle force is given by
\begin{eqnarray}
{F}_{\nu} (\bar{\boldsymbol{p}}) &=\sum_{\mu} M_{\nu\mu} {J}_{\mu} (\bar{\boldsymbol{p}})  + O(\epsilon^2),
\end{eqnarray}
where $M_{\mu \nu}$ is the inverse of the Onsager matrix defined as
\begin{eqnarray}
M_{\nu\mu} &=\sum_{i, j | i>j} S_{j \to i}(\nu) \alpha_{j \to i}  S_{j \to i}(\mu).
\end{eqnarray}
We can check that the Onsager reciprocal relations $M_{\nu\mu} = M_{\mu\nu}$ hold. If we used the fundamental set of cycle~\cite{schnakenberg1976network}, the matrix $M_{\mu \nu}$ is regular, and the relation
\begin{eqnarray}
{J}_{\mu} (\bar{\boldsymbol{p}}) &=\sum_{\nu} M^{-1}_{\mu\nu}  {F}_{\nu} (\bar{\boldsymbol{p}})+ O(\epsilon^2),
\end{eqnarray}
also holds for the inverse matrix $M^{-1}_{\mu\nu}$. The matrix $M^{-1}_{\mu\nu}$ is known as the Onsager matrix. The Onsager reciprocal relations $M^{-1}_{\nu\mu} = M^{-1}_{\mu\nu}$ also hold because of $M_{\nu\mu} = M_{\mu\nu}$.

We start with a generalization of the above discussion for the probability $\boldsymbol{p}$, where $\delta {\boldsymbol{p}}$ is small enough $\delta {\boldsymbol{p}} \simeq O(\epsilon)$.  The excess force is the proportional to the excess flux as follows,
\begin{eqnarray}
\delta {F}_{j \to i}(\boldsymbol{p}|\bar{\boldsymbol{p}}) &=\alpha_{j \to i} \delta {J}_{j \to i}(\boldsymbol{p}|\bar{\boldsymbol{p}}) + O(\epsilon^2),
\end{eqnarray}
where we used Eqs. (\ref{fluxdelta}), (\ref{forcedelta}) and (\ref{coefficientorder}).
If we do not assume the steady-state, the cycle matrix is not regular in general. Thus, we replace the cycle matrix with a regular matrix $\mathcal{S}_{j \to i}(\mu)$. The $\mu$-mode is given by the set of edges, which is not a cycle necessarily.
We defined the $\mu$-th mode of the force and flux as
\begin{eqnarray} 
F_{\mu} ({\boldsymbol{p}})  =\sum_{i, j | i>j} {F}_{j \to i} ({\boldsymbol{p}})  \mathcal{S}_{j \to i}(\mu), \\
J_{j \to i}({\boldsymbol{p}}) =\sum_{\mu} {J}_{\mu} ({\boldsymbol{p}})  \mathcal{S}_{j \to i}(\mu),
\end{eqnarray}
respectively, where the regular matrix $\mathcal{S}_{j \to i}(\mu)$ also satisfies the condition that $\mathcal{S}_{j \to i}(\mu) =-\mathcal{S}_{i \to j}(\mu)=1$ if the $\mu$-th mode includes the edge $j \to i$, and that $\mathcal{S}_{j \to i}(\mu)=\mathcal{S}_{i \to j}(\mu) =0$ if the $\mu$-th mode does not include the edge $j \to i$. 
In a similar way, we define the $\mu$-th mode of the excess force and excess flux as 
\begin{eqnarray} 
\delta F_{\mu} ({\boldsymbol{p}}|| \bar{\boldsymbol{p}})  =\sum_{i, j | i>j} \delta {F}_{j \to i} ({\boldsymbol{p}} ||\bar{\boldsymbol{p}} )  \mathcal{S}_{j \to i}(\mu), \\
\delta J_{j \to i}({\boldsymbol{p}}|| \bar{\boldsymbol{p}}) =\sum_{\mu}\delta J_{\mu} ({\boldsymbol{p}}|| \bar{\boldsymbol{p}})  \mathcal{S}_{j \to i}(\mu).
\end{eqnarray}
 Thus, the $\nu$-th mode of the excess force is given by
\begin{eqnarray}
\delta F_{\nu} ({\boldsymbol{p}}|| \bar{\boldsymbol{p}}) &=\sum_{\mu} \mathcal{M}_{\nu\mu} \delta J_{\mu} ({\boldsymbol{p}}|| \bar{\boldsymbol{p}})  + O(\epsilon^2),
\end{eqnarray}
where $\mathcal{M}_{\nu\mu}$ is the inverse of the Onsager matrix defined as
\begin{eqnarray}
\mathcal{M}_{\nu\mu} &=\sum_{i, j | i>j} \mathcal{S}_{j \to i}(\nu) \alpha_{j \to i}  \mathcal{S}_{j \to i}(\mu).
\end{eqnarray}
We also consider the inverse matrix ${\mathcal{S}^{-1}}_{j \to i}(\mu)$ which satisfies 
\begin{eqnarray} 
 \delta {F}_{j \to i} ({\boldsymbol{p}} ||\bar{\boldsymbol{p}} )  = \sum_{\mu} {\mathcal{S}^{-1}}_{j \to i}(\mu) \delta F_{\mu} ({\boldsymbol{p}}|| \bar{\boldsymbol{p}}) , \\
\delta J_{\mu} ({\boldsymbol{p}}|| \bar{\boldsymbol{p}})  = \sum_{i, j | i>j} {\mathcal{S}^{-1}}_{j \to i}(\mu) \delta J_{j \to i}({\boldsymbol{p}}|| \bar{\boldsymbol{p}}).
\end{eqnarray}
Thus, the $\nu$-th mode of the excess flux is given by
\begin{eqnarray}
\delta J_{\nu} ({\boldsymbol{p}}|| \bar{\boldsymbol{p}}) &=\sum_{\mu} {\mathcal{M}^{-1}}_{\nu\mu} \delta F_{\mu} ({\boldsymbol{p}}|| \bar{\boldsymbol{p}})  + O(\epsilon^2),
\end{eqnarray}
where ${\mathcal{M}^{-1}}_{\nu\mu}$ is the Onsager matrix defined as
\begin{eqnarray}
{\mathcal{M}^{-1}}_{\nu\mu} &=\sum_{i, j | i>j} {\mathcal{S}^{-1}}_{j \to i}(\nu) (\alpha_{j \to i})^{-1}  {\mathcal{S}^{-1}}_{j \to i}(\mu).
\end{eqnarray}
By definition, the Onsager reciprocal relations $\mathcal{M}_{\nu\mu} = \mathcal{M}_{\mu\nu}$ and ${\mathcal{M}^{-1}}_{\nu\mu} = {\mathcal{M}^{-1}}_{\mu\nu}$ holds.

Therefore, the excess entropy production rate is written as the quadratic form of $\delta F_{\nu} ({\boldsymbol{p}}|| \bar{\boldsymbol{p}})$ and $\delta J_{\nu} ({\boldsymbol{p}}|| \bar{\boldsymbol{p}})$ with the inverse of the Onsager matrix $\mathcal{M}_{\nu\mu}$ and the Onsager matrix ${\mathcal{M}^{-1}}_{\nu\mu}$, respectively,
\begin{eqnarray}
\delta^2 \sigma  (\boldsymbol{p} || \bar{\boldsymbol{p}}) &= \sum_{i,j|i>j} \delta F_{i \to j} (\boldsymbol{p} || \bar{\boldsymbol{p}}) \delta J_{i \to j} (\boldsymbol{p} || \bar{\boldsymbol{p}}) \nonumber\\
&=\sum_{\mu} \delta F_{\mu} ({\boldsymbol{p}}|| \bar{\boldsymbol{p}}) \delta J_{\mu} ({\boldsymbol{p}}|| \bar{\boldsymbol{p}}) \nonumber \\
&=\sum_{\mu,\nu} \mathcal{M}_{\nu \mu} \delta J_{\mu} ({\boldsymbol{p}}|| \bar{\boldsymbol{p}}) \delta J_{\nu} ({\boldsymbol{p}}|| \bar{\boldsymbol{p}}) \nonumber \\
&=\sum_{\mu,\nu} {\mathcal{M}^{-1}}_{\nu\mu} \delta F_{\mu} ({\boldsymbol{p}}|| \bar{\boldsymbol{p}}) \delta F_{\nu} ({\boldsymbol{p}}|| \bar{\boldsymbol{p}}).
\label{linearirrevexcessentropy}
\end{eqnarray}
In terms of the Glansdorff-Prigogine criterion for stability, we can discuss the stability of the steady-state $\bar{\boldsymbol{p}}$ based on the Onsager matrix ${\mathcal{M}^{-1}}_{\nu\mu}$ and the inverse of the Onsager matrix $\mathcal{M}_{\nu \mu}$ under the near-equilibrium condition. We remark that the entropy production rate is also given by
\begin{eqnarray}
\sigma  (\bar{\boldsymbol{p}}) &= \sum_{i,j|i>j} F_{i \to j} (\bar{\boldsymbol{p}}) J_{i \to j} (\bar{\boldsymbol{p}}) \nonumber\\
&=\sum_{\mu,\nu} \mathcal{M}_{\nu \mu} J_{\mu}  (\bar{\boldsymbol{p}}) J_{\nu}  (\bar{\boldsymbol{p}}) \nonumber \\
&=\sum_{\mu,\nu} {\mathcal{M}^{-1}}_{\nu\mu}  F_{\mu}  (\bar{\boldsymbol{p}}) F_{\nu}  (\bar{\boldsymbol{p}}).
\end{eqnarray}
Because the second law of thermodynamics $\sigma  (\bar{\boldsymbol{p}}) \geq 0$, the Onsager matrix ${\mathcal{M}^{-1}}_{\nu\mu}$ and the inverse of the Onsager matrix $\mathcal{M}_{\nu \mu}$ are the nonnegative-definite matrices, which satisfy $\sum_{\mu,\nu}  {\mathcal{M}^{-1}}_{\nu\mu} x_{\nu} x_{\mu} \geq 0$ and $\sum_{\mu,\nu}  {\mathcal{M}}_{\nu\mu} x_{\nu} x_{\mu} \geq 0$ for any vector $\{x_{\nu} \}$. Because the Onsager matrix ${\mathcal{M}^{-1}}_{\nu\mu}$ and the inverse of the Onsager matrix $\mathcal{M}_{\nu \mu}$ are the nonnegative-definite matrices, Eq.~(\ref{linearirrevexcessentropy}) implies the stability of the steady-state
\begin{eqnarray}
\delta^2 \sigma  (\boldsymbol{p} || \bar{\boldsymbol{p}}) \geq 0.
\end{eqnarray}

Under the near-equilibrium condition, the Fisher metric is related to the inverse of the Onsager matrix and the Onsager matrix. The quantity $\delta p_i/\bar{p}_i$ is calculated as
\begin{eqnarray}
\frac{\delta p_i}{\bar{p}_i} &= \ln p_i - \ln \bar{p}_i + O(|\delta \boldsymbol{p} |^2) \nonumber\\
&= \sum_{\mu} \left. (\partial_{J_{\mu}} \ln p_i) \right|_{\boldsymbol{p} =\bar{\boldsymbol{p}}}\delta J_{\mu} ({\boldsymbol{p}}|| \bar{\boldsymbol{p}})  + O(|\delta \boldsymbol{p} |^2) \nonumber\\
&= \sum_{\mu} \left. (\partial_{F_{\mu}} \ln p_i) \right|_{\boldsymbol{p} =\bar{\boldsymbol{p}}}\delta F_{\mu} ({\boldsymbol{p}}|| \bar{\boldsymbol{p}})  + O(|\delta \boldsymbol{p} |^2) \nonumber\\
\end{eqnarray}
Thus, the square of the line element around the steady-state $\delta \boldsymbol{p}  = d\boldsymbol{p}$ is given by
\begin{eqnarray}
ds^2 (\bar{\boldsymbol{p} }) &= \sum_{\mu, \nu}g^{J}_{\mu \nu} (\bar{\boldsymbol{p}} ) \delta J_{\mu} ({\boldsymbol{p}}|| \bar{\boldsymbol{p}})  \delta J_{\nu} ({\boldsymbol{p}}|| \bar{\boldsymbol{p}}) + O(|d \boldsymbol{p} |^3), \nonumber \\
g^{J}_{\mu \nu} (\bar{\boldsymbol{p}} )&=\sum_{i} \bar{p}_i \left. (\partial_{J_{\mu}} \ln p_i)\right|_{\boldsymbol{p} =\bar{\boldsymbol{p}}} \left. (\partial_{J_{\nu}} \ln p_i) \right|_{\boldsymbol{p} =\bar{\boldsymbol{p}}} \nonumber \\
&= \mathbb{E}_{\bar{\boldsymbol{p}}} \left[\left. (\partial_{J_{\mu}} \ln p_i)\right|_{\boldsymbol{p} =\bar{\boldsymbol{p}}} \left.( \partial_{J_{\nu}}\ln p )\right|_{\boldsymbol{p} =\bar{\boldsymbol{p}}} \right],
\end{eqnarray}
and
\begin{eqnarray}
ds^2 (\bar{\boldsymbol{p} }) &= \sum_{\mu, \nu}g^{F}_{\mu \nu} (\bar{\boldsymbol{p}} ) \delta F_{\mu} ({\boldsymbol{p}}|| \bar{\boldsymbol{p}})  \delta F_{\nu} ({\boldsymbol{p}}|| \bar{\boldsymbol{p}}) + O(|d \boldsymbol{p} |^3), \nonumber \\
g^{F}_{\mu \nu} (\bar{\boldsymbol{p}} )&=\sum_{i} \bar{p}_i \left. (\partial_{F_{\mu}} \ln p_i)\right|_{\boldsymbol{p} =\bar{\boldsymbol{p}}} \left. (\partial_{F_{\nu}} \ln p_i) \right|_{\boldsymbol{p} =\bar{\boldsymbol{p}}} \nonumber \\
&= \mathbb{E}_{\bar{\boldsymbol{p}}} \left[\left. (\partial_{F_{\mu}} \ln p_i)\right|_{\boldsymbol{p} =\bar{\boldsymbol{p}}} \left.( \partial_{F_{\nu}}\ln p )\right|_{\boldsymbol{p} =\bar{\boldsymbol{p}}} \right].
\end{eqnarray}
Here, $g^{J}_{\mu \nu} (\bar{\boldsymbol{p}})$ and $g^{F}_{\mu \nu} (\bar{\boldsymbol{p}})$ are the Fisher information matrices for the mode of the flux and the force respectively. The matrices $g^{J}_{\mu \nu} (\bar{\boldsymbol{p}})$ and $g^{F}_{\mu \nu} (\bar{\boldsymbol{p}})$ can be interpreted as the metric tensor for the steady-state $\bar{\boldsymbol{p}}$ under the near-equilibrium condition. From the relation between the line element and the excess entropy Eq.~(\ref{relationship}), we obtain information-geometric interpretations of the Onsager matrix and the inverse of the Onsager matrix as follows, 
\begin{eqnarray}
\sum_{\mu,\nu} \mathcal{M}_{\nu \mu} \delta J_{\mu} ({\boldsymbol{p}}|| \bar{\boldsymbol{p}})  \delta J_{\nu} ({\boldsymbol{p}}|| \bar{\boldsymbol{p}}) = -\frac{1}{2} \frac{d}{dt} \left[ \sum_{\mu, \nu}g^{J}_{\mu \nu} (\bar{\boldsymbol{p}} ) \delta J_{\mu} ({\boldsymbol{p}}|| \bar{\boldsymbol{p}})  \delta J_{\nu} ({\boldsymbol{p}}|| \bar{\boldsymbol{p}})  \right], 
\label{onsagerinfogeo1}
\end{eqnarray}
\begin{eqnarray}
\sum_{\mu,\nu} {\mathcal{M}^{-1}}_{\nu\mu} \delta F_{\mu} ({\boldsymbol{p}}|| \bar{\boldsymbol{p}})  \delta F_{\nu} ({\boldsymbol{p}}|| \bar{\boldsymbol{p}}) = -\frac{1}{2} \frac{d}{dt} \left[ \sum_{\mu, \nu}g^{F}_{\mu \nu} (\bar{\boldsymbol{p}} ) \delta F_{\mu} ({\boldsymbol{p}}|| \bar{\boldsymbol{p}})  \delta F_{\nu} ({\boldsymbol{p}}|| \bar{\boldsymbol{p}})  \right]. \nonumber \\
\label{onsagerinfogeo2}
\end{eqnarray}

\section{Disadvantage of the Glansdorff-Prigogine criterion for stability and new criterion for stability based on information geometry}
Based on the above discussion, we discuss a disadvantage of the Glansdorff-Prigogine criterion for stability for the master equation~\cite{Keizer1974criterion, Wilhelm1999what}. For the linear master equation, the Glansdorff-Prigogine criterion for stability is equal to the Lyapunov criterion, but the Glansdorff-Prigogine criterion for stability always claims that the steady-state is stable. On the other hand, the Glansdorff-Prigogine criterion for stability might not be regarded as the Lyapunov criterion for the nonlinear master equation. Thus, the Glansdorff-Prigogine criterion for stability is not so useful for the master equation at least.

\begin{figure}[tb]
\centering
\includegraphics[width=8cm]{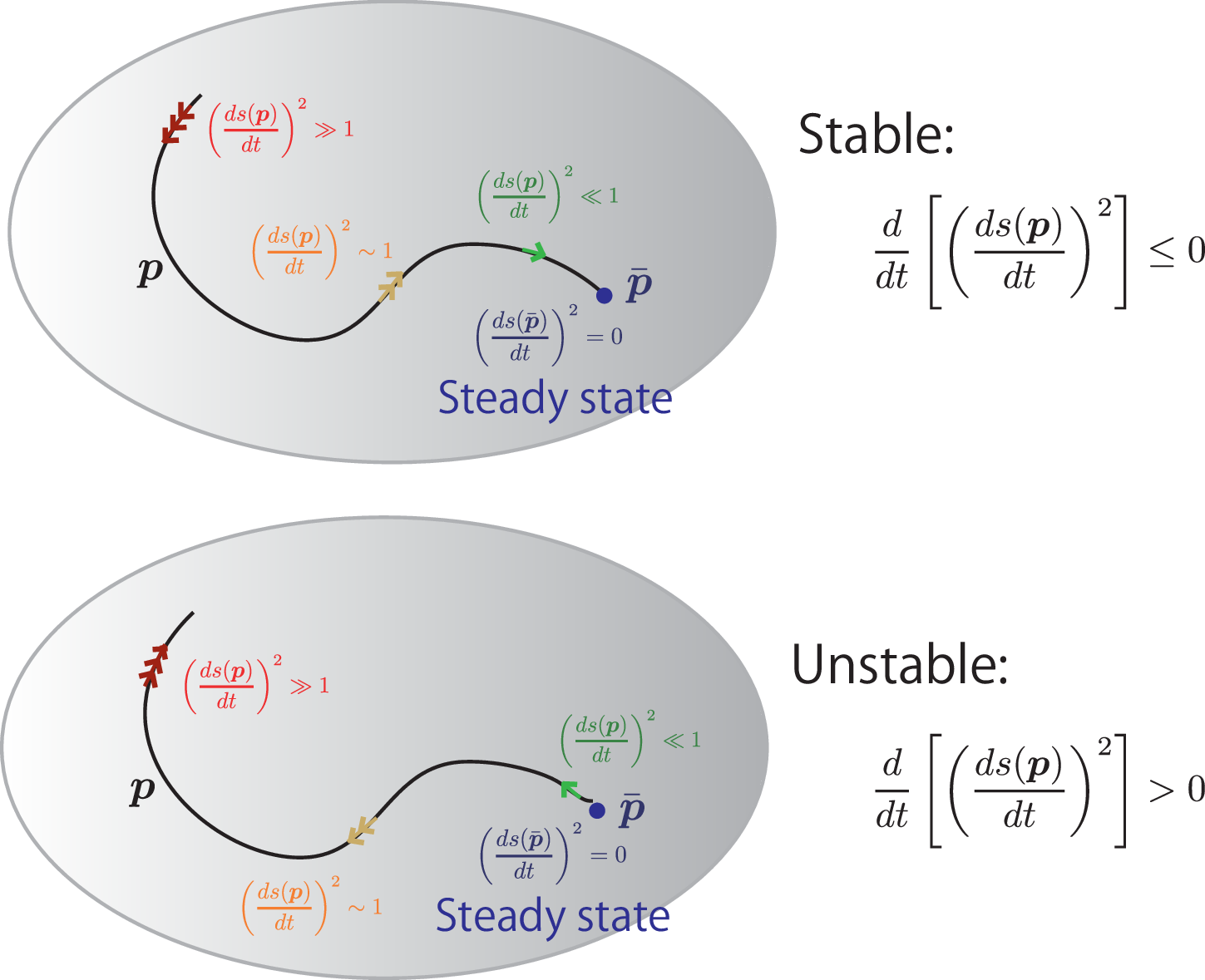}
\caption{Schematic of our information-geometric criterion for stability. Here we show a region of the probability simplex (gray region) that $\delta \boldsymbol{p} =\boldsymbol{p}- \bar{\boldsymbol{p}}$ is small. The black line indicates the path of the time evolution of the probability $\boldsymbol{p}$, and the blue dot indicates the steady-state distribution $\bar{\boldsymbol{p}}$, which is the fixed point on the manifold. In the relaxation process to the stable steady-state, the square of the intrinsic speed $(ds(\boldsymbol{p})/dt)^2$ is decreasing in time. The color of arrow indicates the amount of the intrinsic speed, i.e., red (fast), yellow (middle) and blue (slow). 
Speed decay around the steady-state $\bar{\boldsymbol{p}}$ indicates that the steady-state $\bar{\boldsymbol{p}}$ is stable in our-information-geometric criterion for stability. Around the unstable steady-state, the probability $\boldsymbol{p}$ is removed from the steady state $\bar{\boldsymbol{p}}$ and the intrinsic speed is accelerated. Speed acceleration around the steady-state $\bar{\boldsymbol{p}}$ indicates that the steady-state $\bar{\boldsymbol{p}}$ is unstable in our-information-geometric criterion for stability.}
\label{fig1}
\end{figure}

 We now propose a new Lyapunov criterion based on information geometry, which is a main claim in this article. Based on a relationship between the Lyapunov candidate function and the square of the line element, we generalize the Lyapunov criterion even for the nonlinear master equation. We here introduce the intrinsic speed in information geometry as a Lyapunov candidate function. The square of the intrinsic speed is defined as the Fisher information of time
\begin{eqnarray}
\left(\frac{ds ({\boldsymbol{p}})}{dt} \right)^{2} &= \sum_{\mu, \nu} g_{\mu \nu} \frac{d\theta_{\mu}}{dt} \frac{d\theta_{\nu}}{dt} \nonumber \\
&= \mathbb{E}\left[ \left(\frac{d \ln p}{dt} \right)^2 \right] \nonumber \\
&= \sum_i \frac{1}{p_i} \left(\frac{d p_i}{dt} \right)^2.
\end{eqnarray}
This quantity is nonnegative
\begin{eqnarray}
\left(\frac{ds ({\boldsymbol{p}})}{dt} \right)^{2} \geq 0
\end{eqnarray}
and zero if and only if the system in the steady-state
\begin{eqnarray}
\left(\frac{ds ({\bar{\boldsymbol{p}}})}{dt} \right)^{2} = 0.
\end{eqnarray}
Thus, the square of the intrinsic speed can be a Lyapunov candidate function around the steady-state. This intrinsic speed is well defined even for the nonlinear master equation. We propose a new stability criterion based on information geometry 
\begin{eqnarray}
\frac{d}{dt} \left[\left(\frac{ds({\boldsymbol{p}})}{dt} \right)^{2} \right]=\frac{d}{dt} \left[\left(\frac{ds({\bar{\boldsymbol{p}} + \delta \boldsymbol{p} })}{dt} \right)^{2} \right] \leq 0 \: \:\: \:\: \:\: \:\: \:({\rm stable}),
\end{eqnarray}
\begin{eqnarray}
\frac{d}{dt} \left[\left(\frac{ds({\boldsymbol{p}})}{dt} \right)^{2} \right]=\frac{d}{dt} \left[\left(\frac{ds({\bar{\boldsymbol{p}} + \delta \boldsymbol{p}})}{dt} \right)^{2} \right] > 0 \: \:\: \:\: \: ({\rm unstable}),
\end{eqnarray}
when the probability $\delta \boldsymbol{p}$ is small enough. The steady-state $\bar{\boldsymbol{p}}$ is said to be stable if $d/dt[ds({\bar{\boldsymbol{p}} + \delta \boldsymbol{p} })/dt]^{2} \leq 0$ for small $\delta \boldsymbol{p}$. The steady-state $\bar{\boldsymbol{p}}$ is said to be unstable if $d/dt[ds({\bar{\boldsymbol{p}} + \delta \boldsymbol{p} })/dt]^{2} > 0$ for small $\delta \boldsymbol{p}$. This criterion is concerned with speed decay on the manifold. If the speed on the manifold is reducing in time around the steady-state $\bar{\boldsymbol{p}}$, the steady-state $\bar{\boldsymbol{p}}$ is stable (see also Fig.~\ref{fig1}).

If the transition rate does not depend on time $dW_{i \to j}/dt =0$, we can prove that the Fisher information of time $(ds ({\boldsymbol{p}})/dt)^2$ is monotonically decreasing in time. We here introduce the matrix-valued generator $\mathcal{G}_{ij} = W_{j \to i} - \delta_{ij} \sum_k W_{i \to k}$ which satisfies $d p_i/dt =\sum_j \mathcal{G}_{ij} p_j $. We explicitly calculate $d/dt(ds((\boldsymbol{p})/dt)^2$ as follows~\cite{Ito2018cramerrao},
\begin{eqnarray}
&\frac{d}{dt}\left[\left(\frac{ds({\boldsymbol{p}})}{dt} \right)^{2} \right] \nonumber\\
=&  - \sum_i \left(\frac{d \ln p_i}{dt} \right)^{2} \left(\sum_j \mathcal{G}_{ij} p_j  \right)+  \sum_i 2\left(\frac{d \ln p_i}{dt} \right) \frac{d}{dt}\left(\sum_j \mathcal{G}_{ij} p_j  \right)\nonumber \\
=& \sum_{i,j} 2\left(\frac{d \ln p_i}{dt} \right) \left( \frac{d\mathcal{G}_{ij}}{dt} \right) p_j \nonumber \\
&+\sum_{i,j} \left(\frac{d \ln p_i}{dt} \right) \left( W_{j \to i} - \delta_{ij} \sum_k W_{i \to k}\right) \left[   2 \left(\frac{d \ln p_j}{dt} \right) - \left(\frac{d \ln p_i}{dt} \right) \right] p_j \nonumber \\
=& \sum_{i,j} 2\left(\frac{d \ln p_i}{dt} \right) \left( \frac{d\mathcal{G}_{ij}}{dt} \right) p_j -\sum_{i,j}   \left[  \frac{d }{dt} \ln \left( \frac{p_j}{p_i} \right) \right]^2 W_{j \to i} p_j.
\label{monotonicity}
\end{eqnarray}
Because the second term is always nonpositive
\begin{eqnarray}
-\sum_{i,j}   \left[  \frac{d }{dt} \ln \left( \frac{p_j}{p_i} \right)  \right]^2 W_{j \to i} p_j \leq 0,
\end{eqnarray}
we obtain the inequality
\begin{eqnarray}
\frac{d}{dt}\left[\left(\frac{ds({\boldsymbol{p}})}{dt} \right)^{2} \right] \leq \sum_{i,j} 2\left(\frac{d \ln p_i}{dt} \right) \left( \frac{d\mathcal{G}_{ij}}{dt} \right) p_j.
\label{monotonicityineq}
\end{eqnarray}
If the transition rate does not depend on time $dW_{i \to j}/dt =0$, this inequality implies the monotonicity of the Fisher information
\begin{eqnarray}
\frac{d}{dt}\left[\left(\frac{ds({\boldsymbol{p}})}{dt} \right)^{2} \right] \leq 0.
\end{eqnarray}
This fact implies the stability of the steady-state for the linear master equation.

For the nonlinear master equation, this monotonicity does not hold in general. The transition rate $W_{i \to j}$ generally depends on the probability ${\boldsymbol{p}}$, $d\mathcal{G}_{ij}/dt$ is calculated as
\begin{eqnarray}
\frac{d\mathcal{G}_{ij}}{dt} =\sum_k \left[ \frac{\partial W_{j \to i}}{\partial p_k} -  \delta_{ij} \sum_l \frac{\partial W_{j \to l}}{\partial p_k} \right] \frac{dp_k}{dt}.
\end{eqnarray}
Therefore, the inequality Eq.~(\ref{monotonicityineq}) is given by
\begin{eqnarray}
\frac{d}{dt}\left[\left(\frac{ds({\boldsymbol{p}})}{dt} \right)^{2} \right] &\leq \sum_{i,j} 2\left(\frac{d \ln p_i}{dt} \right) \left( \frac{d\mathcal{G}_{ij}}{dt} \right) p_j \nonumber \\
&= \sum_{i,j,k} 2\left(\frac{d \ln p_i}{dt} -\frac{d \ln p_j}{dt} \right) \left(\frac{\partial W_{j \to i}}{\partial p_k} \frac{dp_k}{dt}\right) p_j.
\end{eqnarray}
Because the right hand side can be positive, the steady-state can be unstable in our information-geometric criterion for stability.

\section{The Cram\'{e}r-Rao inequality and trade-off relations}
Based on the intrinsic speed $ds({\boldsymbol{p}})/dt$, we can obtain a trade-off relation between speed and the fluctuation of the observable. This fact is known as the Cram\'{e}r-Rao inequality~\cite{Rao1992information, Amari2016infogeo} in information theory. Because thermodynamic trade-off relations such as the thermodynamic uncertainty relations is mathematically based on the Cram\'{e}r-Rao inequality for the path probability~\cite{Dechant2018multidimensional,Hasegawa2019uncertainty}, the result in this section can be interpreted as a variant of thermodynamic trade-off relations.

We interpret the Cram\'{e}r-Rao inequality as a trade-off relation. We consider an observable $R_i$ which is a function of the microscopic state $i$. The fluctuation of the observable for the probability ${\boldsymbol{p}}$ is expressed as the standard deviation
\begin{eqnarray}
\sqrt{{\rm Var}_{\boldsymbol{p}}[R]} &= \sqrt{\mathbb{E}_{\boldsymbol{p}}[R^2] -(\mathbb{E}_{\boldsymbol{p}}[R])^2}.
\end{eqnarray}
Let $\tau_R (\boldsymbol{p})$ be the time which satisfies  
\begin{eqnarray}
\tau_R (\boldsymbol{p}) \left| \frac{d}{dt}\mathbb{E}_{\boldsymbol{p}}[R] \right| = \sqrt{{\rm Var}_{\boldsymbol{p}}[R]}.
\end{eqnarray}
It means that the mean values $\mathbb{E}_{\boldsymbol{p}(t)}[R]$ and $\mathbb{E}_{\boldsymbol{p}(t)}[R] + \tau_R (\boldsymbol{p}) \left|d\mathbb{E}_{\boldsymbol{p}}[R] /dt\right| \simeq \mathbb{E}_{\boldsymbol{p}(t+ \tau_R (\boldsymbol{p}(t))}[R]$ can be distinguishable against the fluctuation of the observable $\sqrt{{\rm Var}_{\boldsymbol{p}}[R]}$ after time $\tau_R (\boldsymbol{p})$. Thus, the reciprocal 
\begin{eqnarray}
v_R(\boldsymbol{p}) = \frac{1}{\tau_R (\boldsymbol{p}) }
\end{eqnarray}
implies the speed of the observable $R_i$. The  Cram\'{e}r-Rao inequality indicates that the upper bound on the speed $v_R(\boldsymbol{p})$ is the intrinsic speed $ds(\boldsymbol{p})/ dt := \sqrt{(ds(\boldsymbol{p})/dt)^2}$~\cite{Ito2018cramerrao},
\begin{eqnarray}
v_R(\boldsymbol{p}) \leq \frac{ds({\boldsymbol{p}})}{dt}.
\end{eqnarray}
In terms of a trade-off relations, the Cram\'{e}r-Rao inequality can be rewritten as
\begin{eqnarray}
\left( \frac{d}{dt}\mathbb{E}_{\boldsymbol{p}}[R] \right)^2 \leq  {\rm Var}_{\boldsymbol{p}}[R] \left( \frac{ds({\boldsymbol{p}})}{dt} \right)^2.
\label{tradeoff}
\end{eqnarray}
This is a trade-off relation among the fluctuation of the observable, the mean change of the observable and the square of the intrinsic speed. 

The Cram\'{e}r-Rao inequality provides upper bounds on the excess entropy production rate and the entropy production rate for the relaxation process driven by the linear master equation. If we consider the observable
\begin{eqnarray}
\Delta \ln p_i = \ln p_i - \ln \bar{p}_i,
\end{eqnarray}
the expected value of $\Delta \ln p_i$ is the Kullback-Leibler divergence
\begin{eqnarray}
\mathbb{E}_{\boldsymbol{p}}[\Delta \ln p] = D_{\rm KL}(\boldsymbol{p} ||\bar{\boldsymbol{p}}).
\end{eqnarray}
Thus, the Cram\'{e}r-Rao inequality for the observable $\Delta \ln p_i$ is given by 
\begin{eqnarray}
\left(- \frac{d}{dt} D_{\rm KL}(\boldsymbol{p} ||\bar{\boldsymbol{p}})  \right)^2 \leq  {\rm Var}_{\boldsymbol{p}}[\Delta \ln p] \left( \frac{ds({\boldsymbol{p}})}{dt} \right)^2.
\label{ineqcramer}
\end{eqnarray}
Around the steady-state $\delta \boldsymbol{p} = O(\epsilon)$, the inequality Eq.~(\ref{ineqcramer}) can also be written as 
\begin{eqnarray}
\delta^2 \sigma(\boldsymbol{p} ||\bar{\boldsymbol{p}})  \leq  \sqrt{{\rm Var}_{\boldsymbol{p} }[\Delta \ln p]} \left( \frac{ds (\boldsymbol{p})}{dt} \right).
\label{tradeoffexcessinfogeo}
\end{eqnarray}
up to the order $O(\epsilon^2)$ for the linear master equation. These result implies a novel thermodynamic trade-off relation between the intrinsic speed $ds (\boldsymbol{p})/dt$ and the excess entropy production rate $\delta^2 \sigma(\boldsymbol{p} ||\bar{\boldsymbol{p}})$. This trade-off relation also implies that the excess entropy production rate should be zero $\delta^2 \sigma(\boldsymbol{p} ||\bar{\boldsymbol{p}})=0$ if the intrinsic speed is zero $ds (\boldsymbol{p})/dt = 0$ because the excess entropy production is nonnegative $\delta^2 \sigma(\boldsymbol{p} ||\bar{\boldsymbol{p}}) \geq 0$ for the linear master equation. Thus, the excess entropy production rate $\delta^2 \sigma(\boldsymbol{p} ||\bar{\boldsymbol{p}})$ should be decreasing if $ds (\boldsymbol{p})/dt$ approaches zero around the steady state.
The inequality Eq.~(\ref{ineqcramer}) also gives the upper bound on the entropy production rate in the case of $\bar{\boldsymbol{p}} = {\boldsymbol{p}}^{\rm eq}$,
\begin{eqnarray}
\sigma(\boldsymbol{p}) \leq  \sqrt{{\rm Var}_{\boldsymbol{p}}[\Delta \ln p]} \left( \frac{ds({\boldsymbol{p}})}{dt} \right).
\label{thermodynamictrade-off}
\end{eqnarray}
We remark that a corresponding result of Eq.~(\ref{thermodynamictrade-off}) for chemical thermodynamics has been already derived in Ref.~\cite{Yoshimura2021information}. 

The Cram\'{e}r-Rao inequality provides a physical interpretation for our information-geometric criterion for stability. From the Cram\'{e}r-Rao inequality, the intrinsic speed implies the speed of an optimal observable $R^* :={\rm argmax}_R v_R(\boldsymbol{p})$,
\begin{eqnarray}
 \left( \frac{ds({\boldsymbol{p}})}{dt} \right) = v_{R^*}(\boldsymbol{p}).
\end{eqnarray}
In this sense, our information-geometric criterion for stability means that
\begin{eqnarray}
\frac{d}{dt}\left[\left(\frac{ds({\bar{\boldsymbol{p}} + \delta \boldsymbol{p} } )}{dt} \right)^{2} \right] \leq 0 &\Leftrightarrow \frac{d}{dt} v_{R^*}({\bar{\boldsymbol{p}} + \delta \boldsymbol{p} }) \leq 0 \:\: \:\: \: ({\rm stable}), \label{stabilityobse1}\\
\frac{d}{dt}\left[\left(\frac{ds({\bar{\boldsymbol{p}} + \delta \boldsymbol{p} })}{dt} \right)^{2} \right] > 0 &\Leftrightarrow \frac{d}{dt} v_{R^*}({\bar{\boldsymbol{p}} + \delta \boldsymbol{p} }) > 0 \: \:\: \:\:  ({\rm unstable}),\label{stabilityobse2}
\end{eqnarray}
when $\delta \boldsymbol{p} $ is small.
Thus, the speed of the observable $R^*$ decreases around the steady-state $\bar{\boldsymbol{p}}$ if the steady-state $\bar{\boldsymbol{p}}$ is stable in our information-geometric criterion for stability.

\section{Remarks on our information-geometric criterion for stability} 
At first, we compare our information-geometric criterion for stability and the Lyapunov criterion for the Lyapunov candidate function $\mathcal{L}_{\rm KL} ({\boldsymbol{p}}|| \bar{\boldsymbol{p}})$,
\begin{eqnarray}
\mathcal{L}_{\rm KL} ({\boldsymbol{p}}|| \bar{\boldsymbol{p}}) = D_{\rm KL} ({\boldsymbol{p}}|| \bar{\boldsymbol{p}}),
\end{eqnarray}
which is the Kullback-Leibler divergence between ${\boldsymbol{p}}$ and $\bar{\boldsymbol{p}}$. This Lyapunov candidate function $\mathcal{L}_{\rm KL} ({\boldsymbol{p}}|| \bar{\boldsymbol{p}})$ is nonnegative, and zero if and only if  ${\boldsymbol{p}} = \bar{\boldsymbol{p}}$.  This Lyapunov criterion can work for the nonlinear master equation because we do not use the property of the master equation to define the Lyapunov candidate function $\mathcal{L}_{\rm KL} ({\boldsymbol{p}}|| \bar{\boldsymbol{p}})$. For the linear master equation, this Lyapunov criterion is approximately equal to the Gransdorff-Prigogine criterion for stability around the steady-state because the following relation
\begin{eqnarray}
\delta^2 \sigma ({\boldsymbol{p}}|| \bar{\boldsymbol{p}})= - \frac{d}{dt} \mathcal{L}_{\rm KL} ({\boldsymbol{p}}|| \bar{\boldsymbol{p}}) + O(|\delta {\boldsymbol{p}}|^3),
\end{eqnarray}
holds. We remark that the quantity $-d /dt (\mathcal{L}_{\rm KL} ({\boldsymbol{p}}|| \bar{\boldsymbol{p}}))$ is called as the boundary part of the nonadiabatic entropy production rate~\cite{Esposito2010three} or the Hatano-Sasa excess entropy production rate~\cite{Hatano2001Steadystate} in stochastic thermodynamics. Thus, Eq.~(\ref{ineqcramer}) can be regarded as the trade-off relation between the intrinsic speed and the boundary part of the nonadiabatic entropy production rate. If we consider the situation $\boldsymbol{p}(0) \simeq \bar{\boldsymbol{p}}$ for the unstable fixed point $\bar{\boldsymbol{p}}$, the square of the intrinsic speed is given by
\begin{eqnarray}
2 \frac{\mathcal{L}_{\rm KL} (\boldsymbol{p}(\Delta t) ||\bar{\boldsymbol{p}})}{\Delta t^2} &\simeq & 2 \frac{\mathcal{L}_{\rm KL} (\boldsymbol{p}(\Delta t) ||\boldsymbol{p}(0))}{\Delta t^2}\nonumber \\
&=& 2 \frac{\mathcal{L}_{\rm KL} (\boldsymbol{p}(\Delta t) | |\boldsymbol{p}(\Delta t)-\frac{d {\boldsymbol{p}}(\Delta t)}{dt} \Delta t)}{\Delta t^2} \nonumber \\
&=& \sum_i \frac{1}{\boldsymbol{p}_i(\Delta t)} \left( \frac{d {\boldsymbol{p}_i}(\Delta t)}{dt} \right)^2+ O(\Delta t) \nonumber \\
 &=& \left( \frac{ds(\boldsymbol{p} (\Delta t))}{dt} \right)^2,
\end{eqnarray}
with the small time $\Delta t$. If we consider the situation $\boldsymbol{p}(\tau) \simeq \bar{\boldsymbol{p}}$ for the stable fixed point $\bar{\boldsymbol{p}}$, the square of the intrinsic speed is given by
\begin{eqnarray}
2\frac{\mathcal{L}_{\rm KL} (\boldsymbol{p}(\tau-\Delta t) ||\bar{\boldsymbol{p}})}{\Delta t^2} &\simeq & 2 \frac{\mathcal{L}_{\rm KL} (\boldsymbol{p}(\tau-\Delta t) ||\boldsymbol{p}(\tau))}{\Delta t^2}\nonumber \\
&=&2 \frac{\mathcal{L}_{\rm KL} (\boldsymbol{p}(\tau-\Delta t) | |\boldsymbol{p}(\tau-\Delta t) + \frac{d {\boldsymbol{p}}(\tau-\Delta t)}{dt} \Delta t)}{\Delta t^2} \nonumber \\
&=& \sum_i \frac{1}{\boldsymbol{p}_i(\tau-\Delta t)} \left( \frac{d {\boldsymbol{p}_i}(\tau-\Delta t)}{dt} \right)^2+ O(\Delta t) \nonumber \\
 &=& \left( \frac{ds(\boldsymbol{p} (\tau-\Delta t))}{dt} \right)^2,
\end{eqnarray}
with the small time $\Delta t$. 
Thus, the Lyapunov candidate function of our information-geometric criterion for stability is the Lyapunov candidate function $\mathcal{L}_{\rm KL} ({\boldsymbol{p}}|| \bar{\boldsymbol{p}})$ divided by $\Delta t^2/2$. 

For the nonlinear master equation, our information-geometric criterion for stability might have an advantage compared to the Lyapunov criterion for the Lyapunov candidate function $\mathcal{L}_{\rm KL} ({\boldsymbol{p}}|| \bar{\boldsymbol{p}})$. For the nonlinear master equation, the steady-state solution $\bar{\boldsymbol{p}}$ is not unique. While the Lyapunov candidate function $\mathcal{L}_{\rm KL} ({\boldsymbol{p}}|| \bar{\boldsymbol{p}})$ is a two-point functions of ${\boldsymbol{p}}$ and $\bar{\boldsymbol{p}}$, the square of intrinsic speed $(ds({\boldsymbol{p}}) /dt)^2$ is a one-point function of ${\boldsymbol{p}}$. In the case of the Lyapunov criterion for the Lyapunov candidate function $\mathcal{L}_{\rm KL} ({\boldsymbol{p}}|| \bar{\boldsymbol{p}})$, we need to treat the time derivative of a two-point function $d/dt [\mathcal{L}_{\rm KL} ({\boldsymbol{p}}|| \bar{\boldsymbol{p}})]$ for each steady-state solution $\bar{\boldsymbol{p}}$. On the other hand, we only need to calculate the time derivative of a one-point function $d/dt[(ds({\boldsymbol{p}})/dt)^2 ]$ in the case of our information-geometric criterion for stability. This fact might be an advantage of our information-geometric criterion for stability when the steady-state solution $\bar{\boldsymbol{p}}$ is not uniquely determined. 

We remark the scope of application.  Our information-geometric criterion for stability is based only on the Lyapunov candidate function $(ds(\boldsymbol{p})/dt)^2$. This Lyapunov candidate function $(ds(\boldsymbol{p})/dt)^2$ is physically meaningful based on the Cram\'{e}r-Rao inequality as the trade-off relations. If we can introduce the probability distribution, the Cram\'{e}r-Rao inequality for the Lyapunov candidate function $(ds(\boldsymbol{p})/dt)^2$ does generally hold regardless of the types of stochastic dynamics.
Thus, our information-geometric criterion for stability might work for the general Markov process and the non-Markov process. As shown in the example, we can apply our information-geometric criterion for stability to the deterministic dynamics of the autocatalytic reaction because we can introduce the probability distribution from the conservation law. 

Our information-geometric criterion for stability cannot be simply applied if it is difficult to introduce the probability distribution in dynamics. For example, it might be difficult to apply our information-geometric criterion for stability to fluid dynamics. On the other hand, the original Glansdorff-Prigogine criterion for stability has been discussed for fluid dynamics~\cite{Glansdorff1971stability}. Even though the general approach for dynamics without the probability distribution is unclear, we might generalize our approach for particular deterministic dynamics, such as the deterministic dynamics on chemical reaction networks. On chemical reaction networks, the intrinsic speed and the Cram\'{e}r-Rao inequality can be generalized~\cite{Yoshimura2021information}. Thus we may use the square of the generalized intrinsic speed as the Lyapunov candidate function for the deterministic dynamics on chemical reaction networks.

We discuss the measurability of the intrinsic speed. The intrinsic speed is experimentally measurable if we can see dynamics of the probability distribution $\boldsymbol{p}(t)$. For example, the intrinsic speed can be measured in the single-cell experiment~\cite{Ashida2020information}. On the other hand, the excess entropy production rate is relatively hard to be measured because measurement of the flux and the force $J_{i \to j}(\boldsymbol{p})$ and $F_{i \to j}(\boldsymbol{p})$ at the current state $\boldsymbol{p}$ and the steady flux and the steady force $J_{i \to j}(\bar{\boldsymbol{p}})$ and $F_{i \to j}(\bar{\boldsymbol{p}})$ at the steady-state $\bar{\boldsymbol{p}}$ is needed to obtain the excess entropy production rate experimentally. 

The intrinsic speed $ds({\boldsymbol{p}})/dt$ is also experimentally accessible via the measurement of the expected value of the observable $R_i$, i.e., $\mathbb{E}_{\boldsymbol{p}} [R]$ and $\mathbb{E}_{\boldsymbol{p}} [R^2]$. The lower bound on the intrinsic speed $v_{R}[{\boldsymbol{p}}]$ is obtained from the measurement of $\mathbb{E}_{\boldsymbol{p}} [R]$ and $\mathbb{E}_{\boldsymbol{p}} [R^2]$, and the equality $v_{R^*}[{\boldsymbol{p}}]= ds({\boldsymbol{p}})/dt$ holds for the efficient observable $R^*_i$. Thus, we can discuss the stability of the system based on its fluctuation $\sqrt{{\rm Var}_{\boldsymbol{p}}[R^*]}$ and its change $|d/dt \mathbb{E}_{\boldsymbol{p}}[R^*] |$ in our information-geometric criterion for stability. For example, we only need to measure the ratio between the fluctuation of the efficient observable $\sqrt{{\rm Var}_{\boldsymbol{p}(t) }[R^*]}$ and its change speed of the efficient observable $|d/dt \mathbb{E}_{\boldsymbol{p}(t)}[R^*] |$ around the steady-state $\boldsymbol{p}(t) \simeq \bar{\boldsymbol{p}}$ to discuss the stability of the steady-state. If the speed of the efficient observable 
\begin{eqnarray}
v_{R^*}(\boldsymbol{p}(t))= \frac{\left|\frac{d}{dt} \mathbb{E}_{\boldsymbol{p}(t)}[R^*] \right|}{ \sqrt{{\rm Var}_{\boldsymbol{p}(t) }[R^*]}},
\end{eqnarray}
is decreasing in time, it implies a relaxation to the stable steady-state. Because $\left|d/dt \mathbb{E}_{\boldsymbol{p}(t)}[R^*] \right|$ quantifies the time response of the efficient observable due to the change of the probability $\boldsymbol{p}$, Eqs.~(\ref{stabilityobse1}) and (\ref{stabilityobse2}) can be regarded as a relation between the stability and the response of the efficient observable.

\section{Example of the nonlinear master equation: Autocatalytic reaction} 
We here discuss a simple example to show an advantage of our information-geometric criterion for stability.
To discuss the difference between the Glansdorff-Prigogine criterion for stability and our information-geometric criterion for stability, we consider a single autocatalytic reaction
\begin{eqnarray}
X + Y  {\rightleftharpoons} 2X.
\end{eqnarray}
Let  $k_+$ be the rate constant for the reaction $X + Y  \rightarrow 2X$ and $k_- $ be the rate constant for the reaction $X + Y  \leftarrow 2X$, respectively. 
This single autocatalytic reaction is a simple example of the nonlinear master equation and it has been well discussed in terms of the Glansdorff-Prigogine criterion for stability~\cite{Keizer1974criterion, Glansdorff1974criterion, Wilhelm1999what}. The rate equation of the single autocatalytic reaction is given by
\begin{eqnarray}
\frac{d}{dt}[X] &= k_{+} [X][Y] - k_{-} [X]^2, \nonumber\\
\frac{d}{dt}[Y] &= -k_{+} [X][Y] + k_{-} [X]^2,
\label{autocatalytic}
\end{eqnarray}
where $[X]$ and $[Y]$ are the concentration of the species $X$ and $Y$, respectively. Because we obtain
\begin{eqnarray}
\frac{d}{dt}[X] + \frac{d}{dt}[Y] &= 0,
\end{eqnarray}
the sum $[X]+ [Y]:=N_{\rm tot}$ does not depend on time. This fact is the conservation law for the concentrations.  Based on this constant $N_{\rm tot}$, we can introduce the probability
\begin{eqnarray}
p_X = \frac{[X]}{N_{\rm tot}}, p_Y = \frac{[Y]}{N_{\rm tot}},
\end{eqnarray}
which satisfies $p_X \geq 0$,  $p_Y \geq 0$ and  $p_X + p_Y= 1$. By using the probability $\boldsymbol{p} =(p_X, p_Y)$, the rate equation can be considered as the nonlinear master equation
\begin{eqnarray}
\frac{d}{dt}p_X = - \frac{d}{dt}p_Y= W_{Y \to X} (\boldsymbol{p}) p_Y - W_{X \to Y}(\boldsymbol{p})  p_X,\\
W_{Y \to X}(\boldsymbol{p})  = K_{+} p_X, \: \: \: W_{X \to Y} (\boldsymbol{p}) = K_{-} p_X.
\end{eqnarray}
with $K_{+}=k_+ N_{\rm tot}$ and $K_{-}=k_- N_{\rm tot}$. We obtain two steady-states $\bar{\boldsymbol{p} } = (\bar{p}_X, \bar{p}_Y)$ and $\bar{\boldsymbol{p} }^* = (\bar{p}_X^*, \bar{p}_Y^*)$ for this nonlinear master equation as follows,
\begin{eqnarray}
\bar{p}_X = \frac{K_{+}}{K_{+} +K_{-}}, \: \: \: \bar{p}_Y = \frac{K_{-}}{K_{+} +K_{-}}, \\
\bar{p}_X^* =0, \: \: \: \bar{p}_Y^*=1.
\end{eqnarray} 

\begin{figure}[tb]
\centering
\includegraphics[width=10cm]{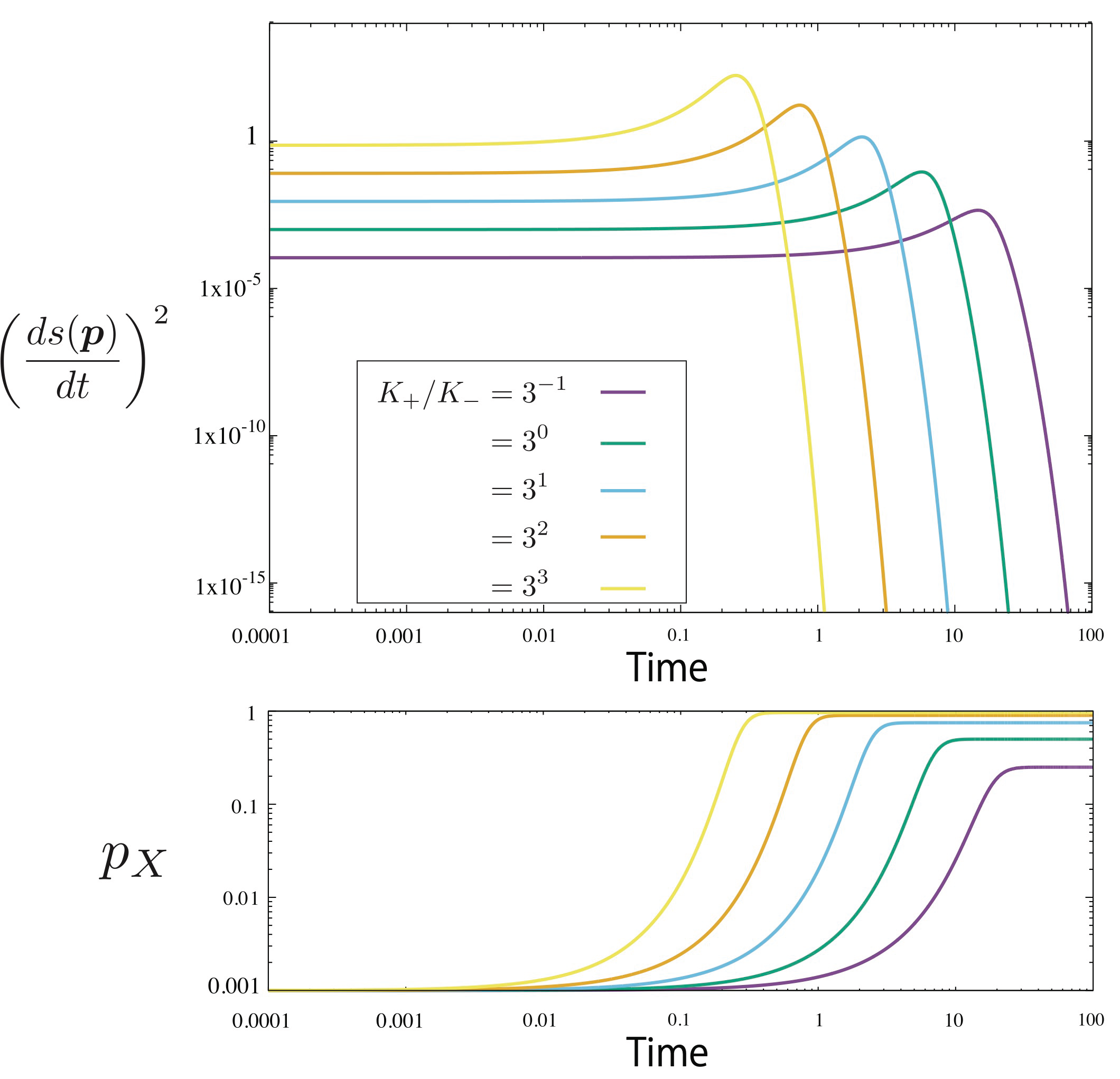}
\caption{The time evolution of the probability $p_X$ and $(ds(\boldsymbol{p})/dt)^2$ for $K_+ =1$. The initial condition is given by $p_X=0.001$, which is around the unstable steady-state $\bar{\boldsymbol{p}}^*$. We can see the behavior of the relaxation to the stable steady-state $\bar{\boldsymbol{p}}$. We shows the time evolution of the square of the intrinsic speed $(ds(\boldsymbol{p})/dt)^2$ and the probability $p_X$ with different choices of parameters $K_+/K_-$. We cannot see any qualitative difference of the steady-state around $K_+/K_- = 3^1$, which is the changeing point from the stability to the unstablity in the Gransdorff-Prigogine criterion for stability. Thus, this numerical calculation implies the disadvantage of the Gransdorff-Prigogine criterion for stability.}
\label{fig2}
\end{figure}
We first discuss that the Glandorff-Prigogine criterion for stability does not work well for the autocatayltic reaction. The excess flux and the excess force for the steady-state $\bar{\boldsymbol{p} }$ can be calculated as
\begin{eqnarray}
\delta J_{Y \to X}(\boldsymbol{p}||\bar{\boldsymbol{p}})&=  [ K_+ (1- 2 \bar{p}_X) +2 K_{-} \bar{p}_X ]\delta p_X, \\
\delta F_{Y \to X}(\boldsymbol{p}||\bar{\boldsymbol{p}}) &= -\left[ \frac{1}{1-\bar{p}_X} + \frac{1}{\bar{p}_X} \right] \delta p_X.
\end{eqnarray}
Then the excess entropy production rate for the steady-state $\bar{\boldsymbol{p}}$ is calculated as
\begin{eqnarray}
\delta^2 \sigma (\boldsymbol{p}||\bar{\boldsymbol{p}}) &= \delta J_{Y \to X} (\boldsymbol{p}||\bar{\boldsymbol{p}})\delta F_{Y \to X} (\boldsymbol{p}||\bar{\boldsymbol{p}})\nonumber \\
&= - \left[\frac{3K_+K_- -K_+^2}{K_+ + K_-} \right]\left[2+ \frac{K_-}{K_+ } +  \frac{K_+}{K_- } \right] (\delta p_X)^2.
\end{eqnarray}
From the viewpoint of the Glandorff-Prigogine criterion for stability, the steady-state $\bar{\boldsymbol{p} }$ can be unstable $\delta^2 \sigma <0$ in the region $3> K_+/ K_-$. But, we cannot see a crucial difference between the region $3> K_+/ K_-$ and  $3 \leq K_+/ K_-$ around the steady-state $\bar{p}_X$ in a numerical simulation (see also Fig.~\ref{fig2}). Indeed, the steady-state $\bar{\boldsymbol{p} }$ is always stable in our information-geometric criterion for stability as discussed later. The excess entropy production rate for the steady-state $\bar{\boldsymbol{p}}^*$ is  also calculated as
\begin{eqnarray}
\delta^2 \sigma (\boldsymbol{p}||\bar{\boldsymbol{p}}^*) &= -K_+ \lim_{\bar{p}^*_X \to +0} \frac{(\delta p_X)^2}{\bar{p}^*_X}.
\end{eqnarray}
Thus, the steady-state $\bar{\boldsymbol{p} }$ is always unstable from the viewpoint of the Glandorff-Prigogine criterion for stability. 

\begin{figure}[tb]
\centering
\includegraphics[width=9cm]{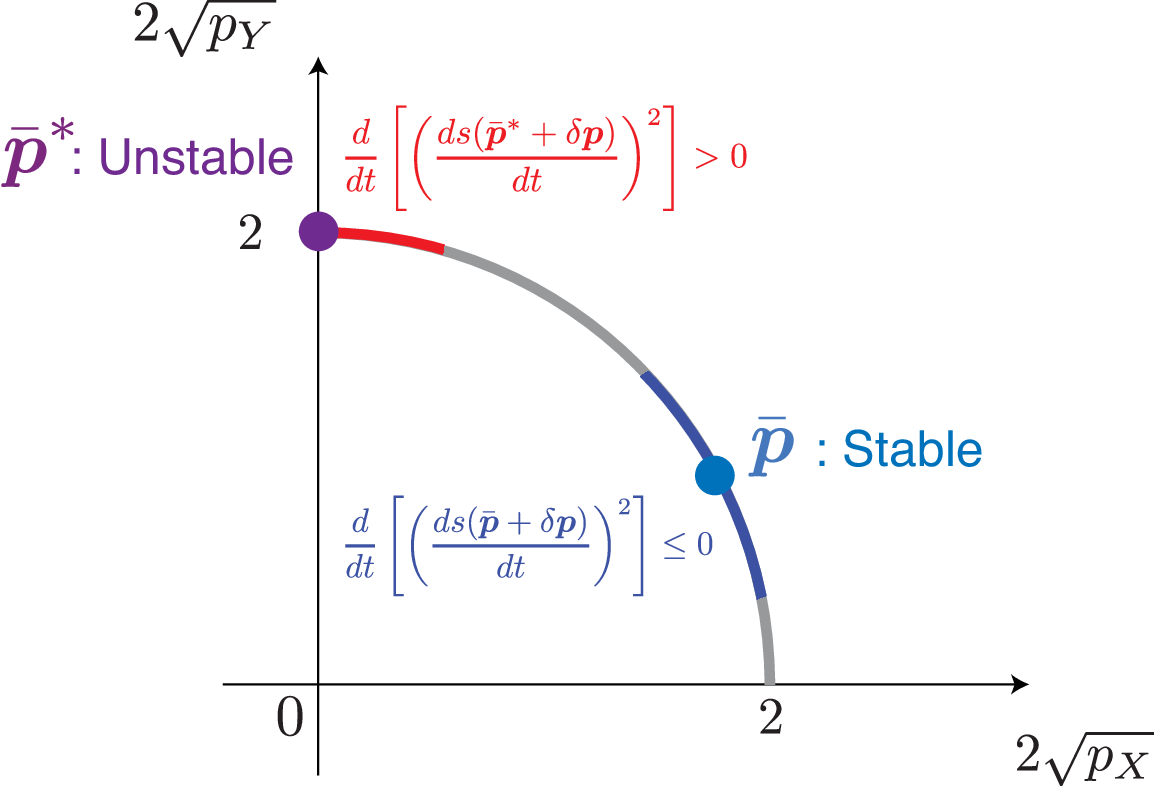} 
\caption{Schematic of information-geometric criterion for stability in the autocatalytic reaction. The circle with the radius $2$ implies the manifold of the probability simplex in information geometry. The steady-state distributions $\bar{\boldsymbol{p}}^*$ and $\bar{\boldsymbol{p}}$ exists on the manifold. The sign of $d/dt[(ds/dt)^2]$ is plus around the unstable steady-state $\bar{\boldsymbol{p}}^*$, and the sign of $d/dt[(ds/dt)^2]$ is minus around the stable steady-state $\bar{\boldsymbol{p}}$.}
\label{fig3}
\end{figure}

In contrast, our information-geometric criterion for stability works well for the autocatayltic reaction. The time derivative of the Fisher information is explicitly calculated as
\begin{eqnarray}
&\frac{d}{dt} \left[ \left(\frac{ds(\boldsymbol{p})}{dt} \right)^{2} \right] \nonumber \\
=& \frac{d}{dt} \left[ \left( \frac{1}{p_Y} + \frac{1}{p_X} \right) \left(\frac{dp_X}{dt} \right)^2 \right] \nonumber \\
=& \left( \frac{1}{p_Y^2} - \frac{1}{p_X^2} \right) \left[K_+ p_X -( K_+ + K_-)p_X^2  \right]^3 \nonumber \\
&+ 2 \left( \frac{1}{p_Y} + \frac{1}{p_X} \right) [ K_+ - 2( K_+ + K_-)p_X] \left[K_+ p_X -( K_+ + K_-)p_X^2  \right]^2.
\end{eqnarray}
Around the steady-state $\bar{\boldsymbol{p}}$, the time derivative of the Fisher information is calculated as
\begin{eqnarray}
\frac{d}{dt} \left[ \left(\frac{ds(\bar{\boldsymbol{p}} + \delta \boldsymbol{p} )}{dt} \right)^{2} \right] = - 2  (K_+)^3 \left[2 + \frac{K_+}{K_-} + \frac{K_-}{K_+} \right] (\delta p_X )^2 + O(|\delta \boldsymbol{p}|^3).
\end{eqnarray}
Therefore,
\begin{eqnarray}
&\frac{d}{dt} \left[ \left(\frac{ds(\bar{\boldsymbol{p}} + \delta \boldsymbol{p} )}{dt} \right)^{2} \right] \leq 0
\end{eqnarray}
when $\delta \boldsymbol{p}$ is small, and the steady-state $\bar{\boldsymbol{p}}$ is always stable in our information-geometric criterion for stability. Around the steady-state $\bar{\boldsymbol{p}}^*$, the time derivative of the Fisher information is also calculated as
\begin{eqnarray}
&\frac{d}{dt} \left[ \left(\frac{ds(\bar{\boldsymbol{p}}^* + \delta \boldsymbol{p} )}{dt} \right)^{2} \right] =(K_+)^3 (\delta p_X ) +O(|\delta \boldsymbol{p}|^2).
\end{eqnarray}
Therefore,
\begin{eqnarray}
&\frac{d}{dt} \left[ \left(\frac{ds(\bar{\boldsymbol{p}}^* + \delta \boldsymbol{p} )}{dt} \right)^{2} \right] > 0
\end{eqnarray}
when $\delta \boldsymbol{p}$ is small, and the steady-state $\bar{\boldsymbol{p}}^*$ is always unstable in our information-geometric criterion for stability. In Fig.~\ref{fig2}, we numerically illustrate the time evolution of the Fisher information $(ds(\boldsymbol{p})/dt)^2$. We show schematic of our information-geometric criterion for stability in Fig.~\ref{fig3}. In the case of the autocatalytic reaction, the square of the line element is given by
\begin{eqnarray}
ds^2({\boldsymbol{p}}) =  (2 d \sqrt{p_X})^2 + (2d \sqrt{p_Y})^2,
\end{eqnarray}
with the constraint $ \sqrt{p_X}^2 +  \sqrt{p_Y}^2 =1$. Thus, we consider a geometry of the circle on the two dimensional Euclidean space with the coordinate $(2\sqrt{p_X}, 2\sqrt{p_Y})$~\cite{Ito2018geometry}. Our information-geometric criterion for stability indicates speed decay on the circle around the steady-state. 

\begin{figure}[tb]
\centering
\includegraphics[width=16cm]{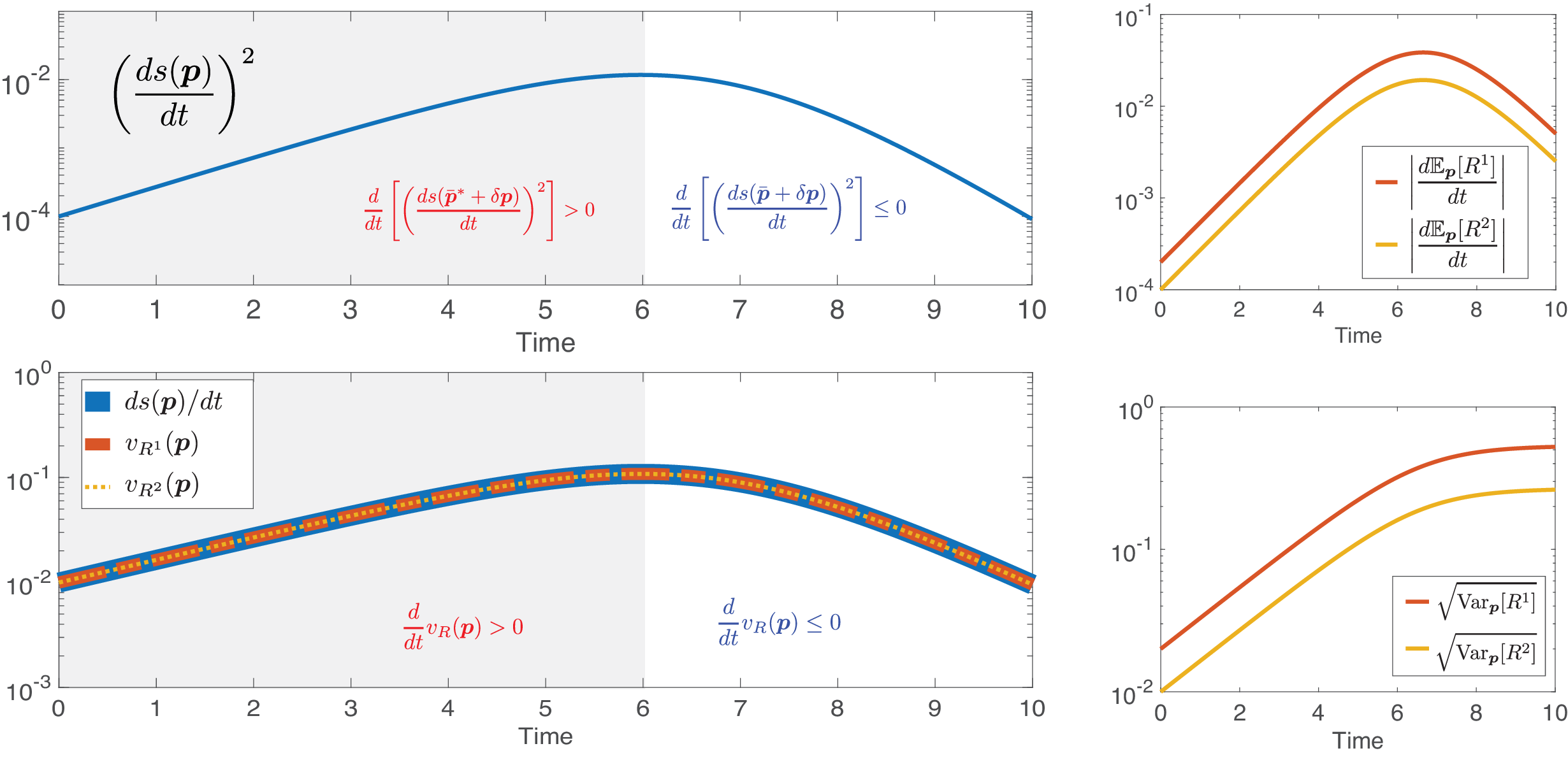}
\caption{The time evolution of the Fisher information $(ds(\boldsymbol{p})/dt)^2$, the intrinsic speed $ds(\boldsymbol{p})/dt$, $|(d/dt) {{\mathbb{E}}_{\boldsymbol{p}}[R]}|$, $\sqrt{{\rm Var}_{\boldsymbol{p}}[R]}$ and the speed of the observable $v_R (\boldsymbol{p}) =|(d/dt) {{\mathbb{E}}_{\boldsymbol{p}}[R]}| / (\sqrt{{\rm Var}_{\boldsymbol{p}}[R]})$. The parameters are given by $K_+=1$ and $K_- =12$, and the initial condition is given by $p_X=0.0001$. We consider two observables $R^1_i$ and $R^2_i$ and numerically check $v_{R^1}  (\boldsymbol{p})=v_{R^2} (\boldsymbol{p})=ds(\boldsymbol{p})/dt$ where $|(d/dt) {{\mathbb{E}}_{\boldsymbol{p}}[R_1]}| \neq |(d/dt) {{\mathbb{E}}_{\boldsymbol{p}}[R_1]}|$ and $\sqrt{{\rm Var}_{\boldsymbol{p}}[R_1]} \neq \sqrt{{\rm Var}_{\boldsymbol{p}}[R_2]}$. The sign of $d/dt(ds(\boldsymbol{p})/dt)^2$ is same as the sign of $dv_{R}(\boldsymbol{p})/dt$. The gray region indicates $dv_{R}(\boldsymbol{p})/dt >0$, and the white region indicates $dv_{R}(\boldsymbol{p})/dt <0$.}
\label{fig4}
\end{figure}
We illustrate the trade-off relation Eq.~(\ref{tradeoff}) by this autocatalytic reaction. We consider the observable $R_i \in \{R_i^1, R_i^2\}$ defined as
\begin{eqnarray}
R^1_i =\delta_{Xi}- \delta_{Yi} = \left\{
\begin{array}{ll}
1 & (i=X), \\
-1 & (i=Y),
\end{array}
\right.
\end{eqnarray}
and 
\begin{eqnarray}
R^2_i =\delta_{Yi} = \left\{
\begin{array}{ll}
0 & (i=X), \\
1 & (i=Y).
\end{array}
\right.
\end{eqnarray}
The expected values of the observable are given by
\begin{eqnarray}
\mathbb{E}_{\boldsymbol{p}}[R^1] = \frac{[X]- [Y]}{[X]+[Y]}, \: \: \: \mathbb{E}_{\boldsymbol{p}}[R^2] = \frac{[Y]}{[X]+[Y]}, 
\end{eqnarray}
and the standard deviations of the observable are given by
\begin{eqnarray}
\sqrt{{\rm Var}_{\boldsymbol{p}}[R^1]} &= \sqrt{1 - \left( \frac{[X]- [Y]}{[X]+[Y]} \right)^2} = \frac{2\sqrt{[X][Y]}}{[X]+[Y]}, \\
\sqrt{{\rm Var}_{\boldsymbol{p}}[R^2]} &= \sqrt{ \frac{[Y]}{[X]+[Y]}-\left( \frac{[Y]}{[X]+[Y]} \right)^2 }  = \frac{\sqrt{[X][Y]}}{[X]+[Y]}.
\end{eqnarray}
We numerically show the time evolution of $|(d/dt)\mathbb{E}_{\boldsymbol{p}}[R]|$ and $\sqrt{{\rm Var}_{\boldsymbol{p}}[R]}$ in Fig.~\ref{fig4}. We also show that the following equality 
\begin{eqnarray}
v_{R^1} (\boldsymbol{p})=v_{R^2} (\boldsymbol{p}) = v_{R^*} (\boldsymbol{p}),
\end{eqnarray}
holds, and thus the trade-off relation $v_{R}(\boldsymbol{p}) \leq ds (\boldsymbol{p})/dt$ holds. Indeed, we can analytically derive the above equality as follows,
\begin{eqnarray}
v_{R}^2 (\boldsymbol{p}) &= \frac{|\frac{d}{dt} \mathbb{E}_{\boldsymbol{p}}[R]|^2}{{\rm Var}_{\boldsymbol{p}}[R]} \nonumber \\
&= \frac{(R_X -R_Y)^2\left(\frac{dp_X}{dt} \right)^2}{R_X^2 p_X + R_Y^2 (1- p_X) - (R_X  p_X + (1-p_X) R_Y)^2}  \nonumber\\
&= \left(\frac{1}{p_X} + \frac{1}{1-p_X} \right)\left(\frac{dp_X}{dt} \right)^2 \nonumber \\
&= \left( \frac{ds(\boldsymbol{p})}{dt}\right)^2.
\end{eqnarray}
Thus, any observable $R_i$ is efficient $v_{R}= v_{R^*}$ for the two dimensional case.
This fact implies that $dv_{R}/dt \leq 0$ in the region $\frac{d}{dt}\left[\left(ds({\boldsymbol{p}})/dt \right)^{2} \right] \leq 0$ around the stable steady-state $\bar{\boldsymbol{p}}$ and $dv_{R}/dt > 0$ in the region $\frac{d}{dt}\left[\left(ds({\boldsymbol{p}})/dt \right)^{2} \right] > 0$ around the unstable steady-state $\bar{\boldsymbol{p}}^*$. Therefore, our information-geometric criterion for stability is meaningful because this trade-off relation gives a behavior of the observable, while the Glandorff-Prigogine criterion for the stability gives an elusive conclusion.

\section{Conclusion} We find a relation between the excess entropy production rate and the square of the line element in information geometry. Around the steady-state $\bar{\boldsymbol{p}}$, the Lyapunov candidate function of the Glansdorff-Prigogine criterion for stability $\delta^2 \mathcal{L}(\boldsymbol{p}||\bar{\boldsymbol{p}})$ implies the square of the line element $ds^2 (\bar{\boldsymbol{p}})$ in information geometry.
Based on this relation, we generalize the Glansdorff-Prigogine criterion for stability. Our generalized criterion is more reasonable than the original criterion because our criterion can be the Lyapunov criterion for the Lyapunov candidate function $(ds (\delta \boldsymbol{p}+\bar{\boldsymbol{p}})/dt)^2$ which works well even for the nonlinear master equation. Thus, our information-geometric criterion for stability is useful to detect the stability of the steady-state in the nonlinear master equation. Our criterion is related to the decay speed $d/dt(ds (\delta \boldsymbol{p}+\bar{\boldsymbol{p}})/dt)^2$ on the manifold of the probability simplex, and naturally understandable from the viewpoint of information geometry. Moreover, a trade-off relation between the intrinsic speed $ds ( \boldsymbol{p})/dt$ and the speed of observable $v_{R}(\boldsymbol{p})$ provides a novel thermodynamic trade-off relation between the intrinsic speed and the excess entropy production rate and a physical interpretation of our information-geometric criterion for stability. 

Our result gives a new information-geometric perspective on the nonlinear master equation. Thermodynamics for nonlinear dynamics has been recently discussed in terms of stochastic thermodynamics and chemical thermodynamics~\cite{Yoshimura2021information,ge2016nonequilibrium,rao2016nonequilibrium, Korbel2021stochastic, Yoshimura2021thermodynamic}. Our result might be useful to understand the stability of such nonlinear dynamics. It might be interesting to consider the bifurcation theory for nonlinear dynamics form the viewpoint of our information-geometric criterion for stability. Because we can discuss the critical phenomena based on the bifurcation theory in statistical physics, our approach might give new information-geometric perspective to the critical phenomena.

\section*{Acknowledgement}
S. I. thank Kohei Yoshimura, Andreas Dechant and Ohga Naruo for fruitful discussions. S. I. also thank Kiyoshi Kanazawa for his careful reading of the manuscript.
S. I. is supported by JSPS KAKENHI Grant No. 19H05796, 21H01560, JST Presto Grant No. JPMJPR18M2 and UTEC-UTokyo FSI Research Grant Program.

\end{document}